\documentclass[aps,prc,twocolumn,superscriptaddress,showpacs,floatfix]{revtex4}

\usepackage{graphicx}
\usepackage{longtable}

\begin{document}

\sloppy

\title{Low-energy dipole strength in $^{112,120}$Sn}

\author{B.~\surname{\"{O}zel-Tashenov}}
\affiliation{Institut f\"ur Kernphysik, Technische Universit\"at Darmstadt, D-64289 Darmstadt, Germany}
\affiliation{Faculty of Science and Letters, Nigde University, 05005 Nigde, Turkey}
\author{J.~\surname{Enders}}
\affiliation{Institut f\"ur Kernphysik, Technische Universit\"at Darmstadt, D-64289 Darmstadt, Germany}
\author{H.~\surname{Lenske}}
\affiliation{Institut f\"ur Theoretische Physik, Justus-Liebig Universit\"{a}t Gie\ss en, D-35392 Gie\ss en, Germany}
\author{A.~M.~\surname{Krumbholz}}
\affiliation{Institut f\"ur Kernphysik, Technische Universit\"at Darmstadt, D-64289 Darmstadt, Germany}
\author{E.~\surname{Litvinova}}
\affiliation{Department of Physics, Western Michigan University, Kalamazoo, MI 49008-5252, USA}
\affiliation{NSCL, Michigan State University, MI 48824, USA}
\author{P.~\surname{von~Neumann-Cosel}}
\email{Email:vnc@ikp.tu-darmstadt.de}
\affiliation{Institut f\"ur Kernphysik, Technische Universit\"at Darmstadt, D-64289 Darmstadt, Germany}
\author{I.~\surname{Poltoratska}}
\affiliation{Institut f\"ur Kernphysik, Technische Universit\"at Darmstadt, D-64289 Darmstadt, Germany}
%\author{V.~Yu.~\surname{Ponomarev}} 
%\affiliation{Institut f\"ur Kernphysik, Technische Universit\"at Darmstadt, D-64289 Darmstadt, Germany}
\author{A.~\surname{Richter}}
\affiliation{Institut f\"ur Kernphysik, Technische Universit\"at Darmstadt, D-64289 Darmstadt, Germany}
\author{G.~\surname{Rusev}}
%\affiliation{Triangle Universities Nuclear Laboratory and Department of Physics, Duke University, Durham, North Carolina 27708, USA}
\affiliation{Chemistry Division, Los Alamos National Laboratory, Los Alamos, New Mexico 87545, USA}
\author{D.~\surname{Savran}}
\affiliation{ExtreMe Matter Institute and Research Devision, GSI, D-64291 Darmstadt, Germany}
\affiliation{Frankfurt Institute for Advanced Studies FIAS, D-60438 Frankfurt am Main, Germany}
\author{N.~\surname{Tsoneva}}
\affiliation{Institut f\"ur Theoretische Physik, Justus-Liebig Universit\"{a}t Gie\ss en, D-35392 Gie\ss en, Germany}
\affiliation{Institute of Nuclear Research and Nuclear Energy, 1784 Sofia, Bulgaria}

\date{\today}

\begin{abstract}

The $^{112,120}$Sn($\gamma,\gamma^{\prime}$) reactions below the neutron separation energies have been studied at the superconducting Darmstadt electron linear accelerator S-DALINAC for different endpoint energies of the incident bremsstrahlung spectrum. 
Dipole strength distributions are extracted for $^{112}$Sn up to 9.5 MeV and for $^{120}$Sn up to 9.1 MeV. 
A concentration of dipole excitations is observed between 5 and 8 MeV in both nuclei. 
Missing strength due to unobserved decays to excited states is estimated in a statistical model. 
A fluctuation analysis is applied to the photon scattering spectra to extract the amount of the unresolved strength hidden in background due to fragmentation. 
The strength distributions are discussed within different model approaches such as the quasiparticle-phonon model and the relativistic time blocking approximation allowing for an inclusion of complex configurations beyond the initial particle-hole states. 
While a satisfactory description of the fragmentation can be achieved for sufficently large model spaces, the predicted centroids and total electric dipole strengths for stable tin isotopes strongly depend on the assumptions about the underlying mean field.

\end{abstract}

\pacs{25.20.Dc, 21.60.Jz, 27.60.+j}

\maketitle

\section{Introduction}

The electric Pygmy Dipole Resonance (PDR) in nuclei is a subject of high current interest (for a recent review, see~\cite{sav13}). 
It is expected to occur at energies well below the IsoVector Giant Dipole Resonance (IVGDR) and may exhaust a considerable fraction of the total electric dipole ($E1$) strength in nuclei with a very asymmetric proton-to-neutron ratio. 
The properties of the mode are claimed to provide insight into the formation of a neutron skin \cite{pie06,kli07,tso08,pie11,ina11}, although this is still under debate \cite{rei10,rei13}. 
It may also constrain the density dependence of the symmetry energy \cite{kli07,car10,fat12,tsa12}. 
Thus, investigations of the PDR will be an important topic at future rare isotope beam facilities. 
Furthermore, dipole strength in the vicinity of the neutron thresholds may lead to significant changes of neutron-capture rates in the astrophysical $r$-process \cite{gor04,lit09,dao12}.

Originally considered to be a single-particle effect \cite{lan71}, many microscopic models nowadays favor an explanation of the PDR as an oscillation of a neutron skin - emerging with an increasing $N/Z$ ratio - against an approximately isospin-saturated core. 
This conclusion is based on the analysis of theoretical transition densities, which differ significantly from those in the IVGDR region.
However, at least for stable nuclei with a moderate neutron excess this question is far from settled, see e.g.\ the recent work of Ref.~\cite{pap14}. 
Quantitative predictions of the centroid energy and strength of the PDR and the corresponding collectivity  as a function of neutron excess differ considerably.
This is partly due to the properties of the underlying mean-field description (e.g., Skyrme-type or relativistic models) and partly results from the unclear separation between PDR and GDR.
$E1$ strength distributions at low excitation energy are also strongly modified in models allowing for complex configuration beyond the 1 particle - 1 hole ($1p1h$) level (see e.g.\ Refs.~\cite{rye02,ton10,lit10}). 

Data on the low-energy $E1$ strength in very neutron-rich heavy nuclei are scarce \cite{adr05,kli07,wie09,ros13}.
Although the PDR strength is much weaker in stable nuclei, detailed spectroscopy with different isovector \cite{kne96,tam11} and isoscalar \cite{pol92,sav06,end10} probes provides important insight into a possible interpretation of the mode as a neutron-skin oscillation, the interplay of collectivity and single-particle degrees of freedom and its isospin nature \cite{paa09,roc12,vre12,yuk12,lan14}.
Extensive studies have been performed in stable even-mass nuclides utilizing the $(\gamma,\gamma')$ reaction, in particular at the shell closures $Z = 20$ \cite{har00,isa11}, $N = 50$ \cite{sch07,sch08,sch13}, $Z = 50$ \cite{gov98}, $N = 82$ \cite{zil02,vol06,sav08,ton10} and in $^{208}$Pb \cite{rye02,sch10}.
However, the connection of these results to the PDR in nuclei with very large $N/Z$ ratios is not clear \cite{sav13,paa07}.

In this respect, a systematic investigation of the PDR in the tin isotope chain is of special interest because a wide range of  isotopes is experimentally accessible while the underlying structure changes only moderatly.
Pioneering experiments on the $E1$ response below the IVGDR in the exotic isotopes $^{130,132}$Sn  and its odd-mass neighbors have been reported \cite{kli07,adr05}.
Very recently, a new experiment has been performed at GSI aiming at an extraction of the complete $E1$ response in $^{124-134}$Sn \cite{aum12}.   
If combined with results on the stable isotopes, for the first time a set of data spanning a large range of $N/Z$ ratios from about 1.25 to 1.68 would be available, which can serve as a benchmark test for the validity of various theoretical approaches. 
Indeed, the Sn isotopes have been a favorite case in the model calculations studying features of the PDR as a function of neutron excess \cite{tso08,pap14,pie06,lit10,sar04,tso04a,tso04b,vre04,paa05,kam06,ter06,lit08,lit13,lan09,co09,avd11,dao11}.

Experimental information on the low-energy $E1$ strength in $^{116}$Sn and $^{124}$Sn is available from Ref.~\cite{gov98}. 
Here we report results from new $(\gamma,\gamma')$ experiments on $^{112}$Sn and $^{120}$Sn which allow to establish systematics of the low-energy $E1$ strength over the range of stable even-mass tin isotopes.  
Beyond the analysis of resolved transitions, in the present work a fluctuation analysis is applied to the ($\gamma,\gamma^{\prime}$) spectra to investigate the amount of unresolved strength which might be hidden in the background because of the fragmentation due to the high level density.
We also estimate in a statistical model approach \cite{rus08} the possible magnitude of corrections to the B($E1$) strengths deduced from the experiments due to unobserved decays to excited states.

The paper is organized as follows:
In Sec.~\ref{sec:experiment} details of the experiment, data analysis and experimental results are given.
Section~\ref{sec:fluc} discusses the extraction of unresolved strength with a fluctuation analysis.
A comparison to theoretical predictions is presented in Sec.~\ref{sec:comp}, and conclusions and an outlook (Sec.~\ref{sec:conclusion}) close the paper.  

%%%%%%%%%%%%%%%%%%%%%%%%%%%%%%%%%%%%%%%%%%%%%%%%%%%%%%%%%%%%%%%%%%%%%%%%%%%%
\section{Experiment and Analysis of Resolved Transitions}
\label{sec:experiment}

\subsection{Experimental details}

Measurements of the $^{112,120}$Sn$(\gamma,\gamma')$ reactions have been performed at the superconducting electron linear accelerator S-DALINAC at the TU Darmstadt. 
Nuclear resonance fluorescence (NRF) is a well-suited tool to investigate low-lying dipole excitations in nuclei and to provide detailed spectroscopic information \cite{kne96}.
The principle of this method is to produce continuous bremsstrahlung spectra of real photons, extending up to the incident electron energy $E_{0}$, which are then used to irradiate the target and simultaneously excite all transitions with a large decay width $\Gamma_0$  into the ground state (g.s.) in the given photon energy range.
Details of the experiment and the data analysis can be found in Ref.~\cite{oze08}.

Experiments on $^{112}$Sn were performed at electron energies of 5.5, 7.0, and 9.5 MeV, while data for $^{120}$Sn were taken at 7.5 and 9.1 MeV, respectively.
The use of different endpoint energies allows to investigate the problem of feeding by higher-lying states, as discussed in Sec.~\ref{subsec:feeding}.
The multipolarities of the transitions were determined from two high purity germanium (HPGe) detectors placed at $90^{\circ}$ and $130^{\circ}$ with respect to incoming photon beam. 
Details of the experimental setup are described in Ref.~\cite{moh99}.
\begin{figure}[tbh!]
    \includegraphics[angle=0,width=8.5cm]{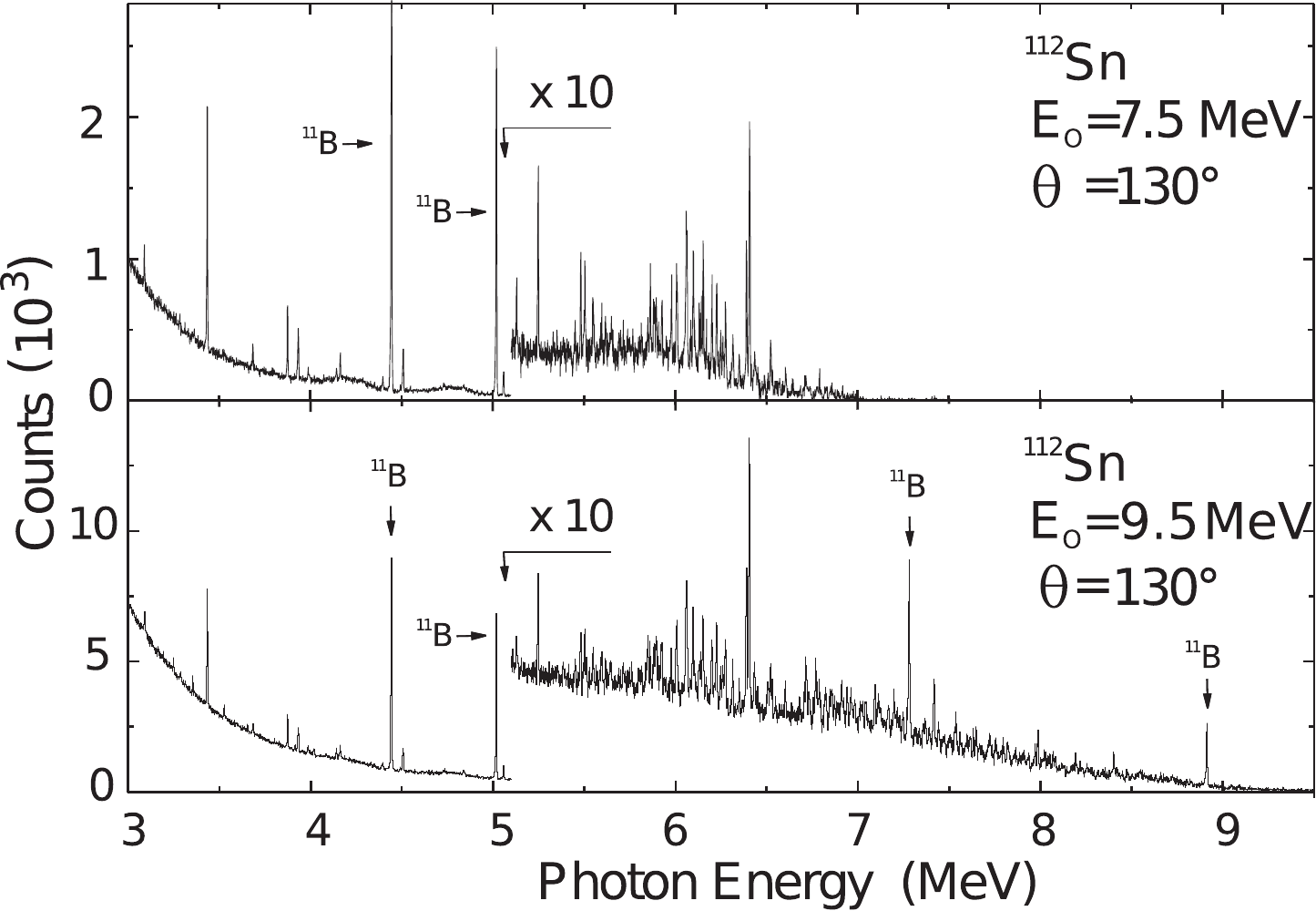}
        \caption{\label{fig:spec1}
Spectra of the $^{112}$Sn$(\gamma,\gamma')$ reaction at endpoint energies of 7.0 MeV (top) and 9.5 MeV (bottom) and a scattering angle $\Theta = 130^\circ$.
Below $E_{\rm x} = 5$ MeV the spectra are suppressed by a factor of 10 because of strongly rising background towards lower energies.
Transitions belonging to the $^{11}$B calibration standard are labeled.}
\end{figure}
\begin{figure}[tbh!]
    \includegraphics[angle=0,width=8.5cm]{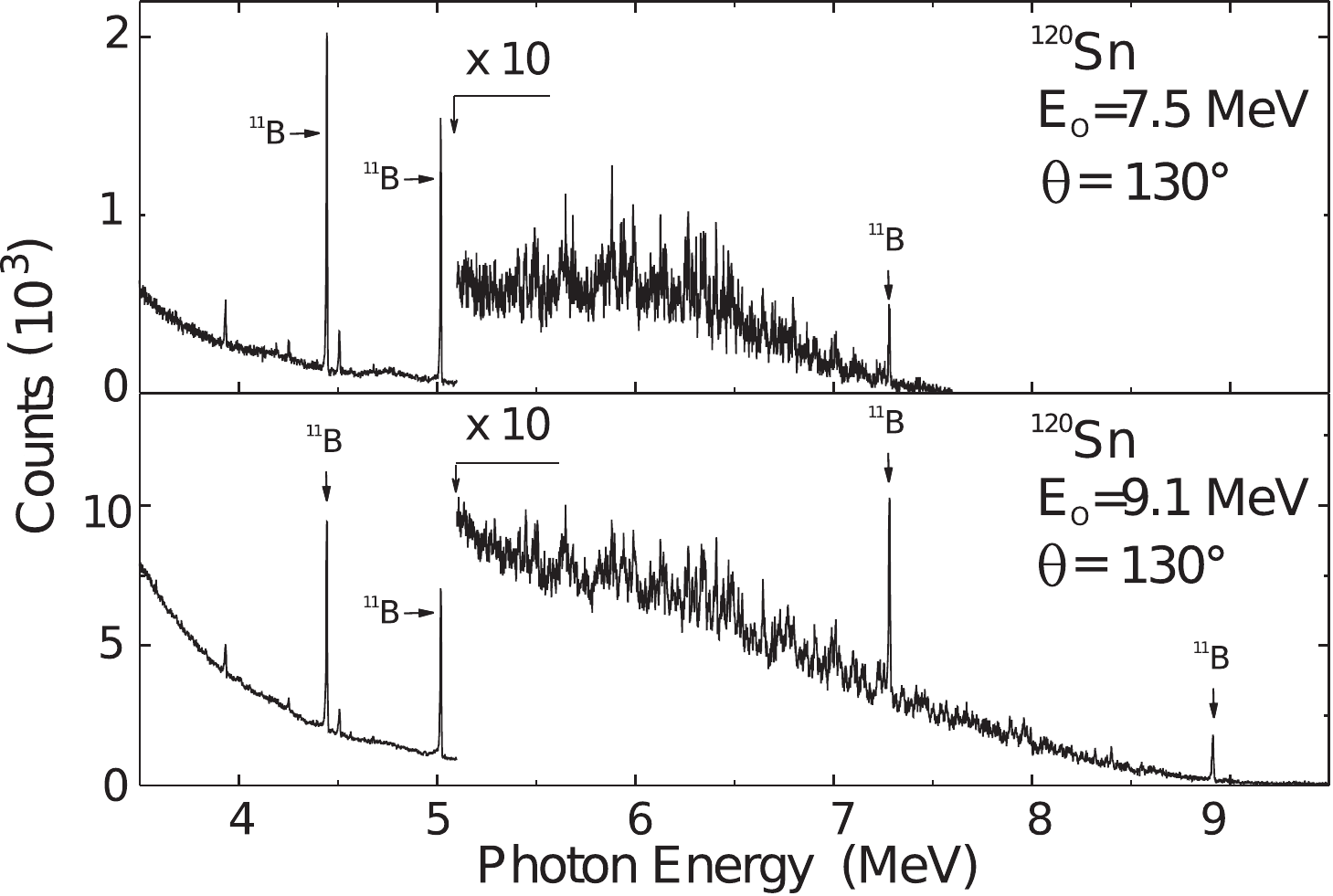}
        \caption{\label{fig:spec2}
Spectra of the $^{120}$Sn$(\gamma,\gamma')$ reaction at endpoint energies of 7.5 MeV (top) and 9.1 MeV (bottom) and a scattering angle $\Theta = 130^\circ$.
Below $E_{\rm x} = 5$ MeV the spectra are scaled by a factor of 10 because of strongly rising background towards lower energies.
Transitions belonging to the $^{11}$B calibration standard are labeled.}
\end{figure}

Targets were made of isotopically enriched ($>$99\%) metallic $^{112}$Sn and $^{120}$Sn samples of about 2 g.
These were sandwiched between $^{11}$B layers with a total weight of about 1~g serving as a standard for energy calibration and the determination of the photon flux and efficiency.
Spectra measured at $\Theta = 130^\circ$ for $^{112}$Sn at 7.0 and 9.5 MeV endpoint energies  and for $^{120}$Sn with 7.5 MeV and 9.1 MeV endpoint energies are displayed in in Figs.~\ref{fig:spec1} and \ref{fig:spec2}, respectively.
The spectra are scaled by a factor of 10 below $E_{\rm x} = 5$ MeV because of the non-resonant background due to atomic processes strongly rising towards lower energies.
Many transitions attributed to the isotpe under investigation are visible in both targets in the region $E_{\rm x} = 5 - 7$ MeV indicating a resonance-like structure.
Overall, the data taken on $^{120}$Sn show more fragmentation than observed in $^{112}$Sn and also compared to the data in $^{116,124}$Sn \cite{gov98}.  

\subsection{Spin determination}

\begin{figure}[t]
    \includegraphics[angle=0,width=7cm]{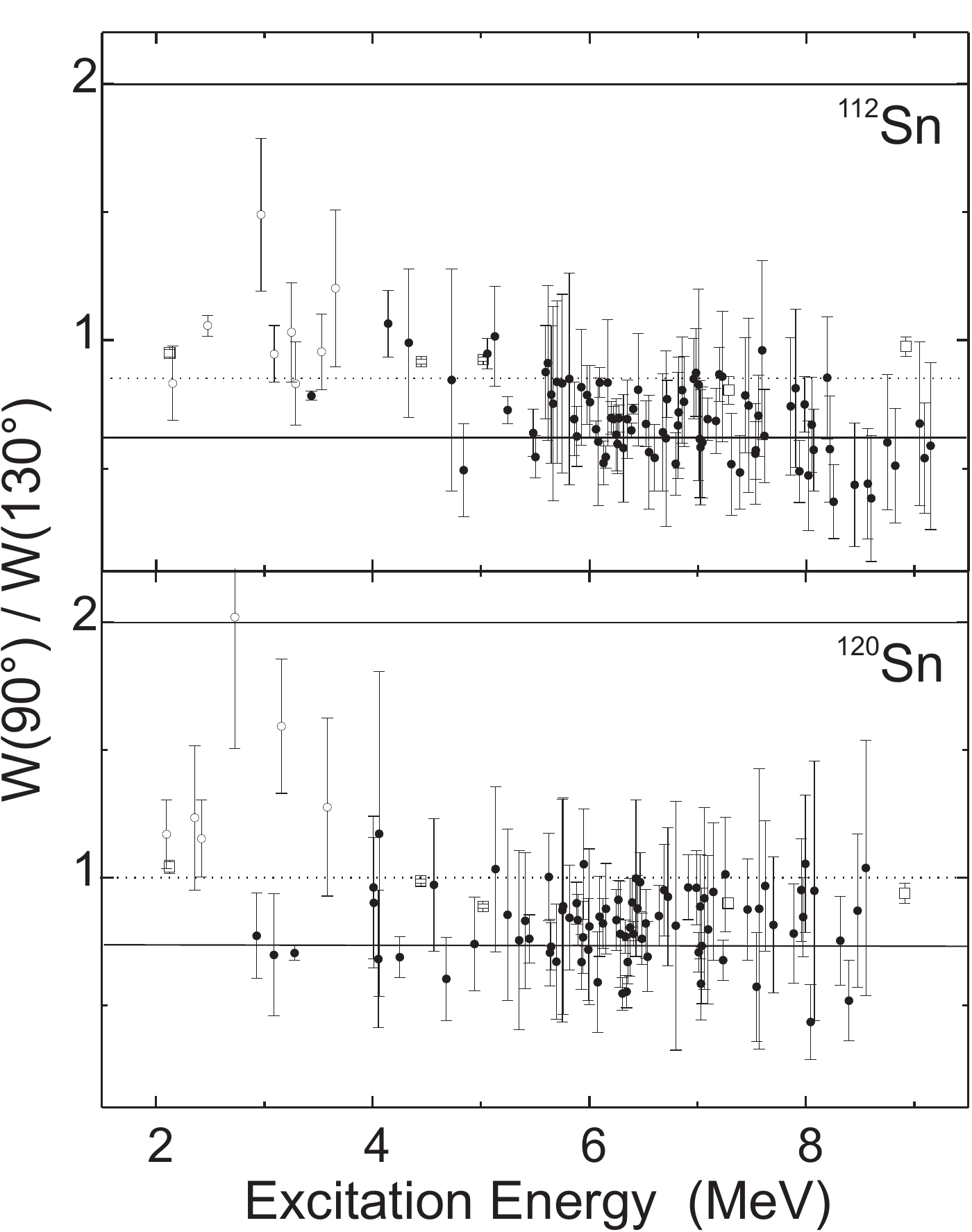}
        \caption{\label{fig:angdist}
Intensity ratios $W(90^{\circ})/(130^{\circ})$ of ground-state transitions observed in $^{112}$Sn (top) and $^{120}$Sn (bottom) for the highest endpoint energies.  
The solid lines are expected values for transition multipolarities $\lambda =1$ (ratio 0.7) or 2 (ratio 2), and the dotted line corresponds to isotropic decay.}
\end{figure}

The spin of the excited states can be determined by comparing the intensities of a given line measured simultanously at different scattering angles. 
Figure~\ref{fig:angdist} shows the ratios measured at 90$^{\circ}$ and 130$^{\circ}$ for both isotopes. 
The solid lines are the values expected for $J = 1$  (0.7) and $J = 2$ states (2.0) starting from a $J = 0$ ground state.. 
The dotted line indicates an isotropic distribution. 
The open squares correspond to $^{11}$B transitions. 
They should be close to the isotropic line because of the half-integer gound state (g.s.) spin which limits deviations of $W(90^\circ)/W(130^\circ)$ from unity to about 10\%, and indeed values close to one are observed. 
Open circles correspond to the excitation of known \cite{nndc} $J^\pi = 2^+$ states and full circles to states with unknown spin except for the quadrupole-octupole two-phonon $1^-$ states \cite{bry99,pys06}.
We note in passing that that for the assumed two-phonon $1^-$ state in $^{112}$Sn, the four-fold segmentation of the $90^\circ$ detector \cite{hut02} was used to extract the multipole character of the transition.
While negative parity is clearly favored, the statistics remains insufficient to exclude positive parity on the $2 \sigma$ confidence level \cite{sie06}.  
  corresponding to $J = 1$ of the excited state.

Known quadrupole transitions in $^{112}$Sn and $^{120}$Sn below 4 MeV populating $2^+$ states deviate from the expected ratio because of feeding.
The intensity ratio for all other ground state transitions is compatible with a dipole character of the transition 
\subsection{Extraction of reduced transition strengths}
\label{subsec:BE1}
Integrated cross sections $I_0$ for photoexcitation of nuclear levels and subsequent decay into the g.s.\ can be derived from the spectra.
They are related to the decay width $\Gamma_0$ into the g.s.\ by
\begin{equation}
    \label{eq:intc}
   I_{0}=\left(\frac{\pi \hbar c}{E_{\rm x}^2} \right)  g \Gamma_0 \frac{\Gamma_{0}}{\Gamma},
\end{equation}
where $\Gamma$ denotes the total width.
$I_0$ is determined from the experimental quantities using the relation
\begin{equation}
    \label{eq:pf}
I_{s}^{i}  = \frac{A_{i}}{N_{T} N_{\gamma}(E_{\rm x},E_{0}) \varepsilon_{abs}(E_{\rm x}) W_{eff}^{i}(\theta)}.
\end{equation}
Here, $A_{i}$ is the peak area of the $i$-th line in the spectrum, $N_T$ denotes the number of target atoms, $W_{eff}^{i}(\theta)$ stands for the effective angular correlation function taking into account the averaging of the finite opening angle of the detectors, and $g=(2J+1)/(2J_0+1)$ is a spin statististical factor averaging over substates of ground state spin $J_0$ and summing over substates of final spin $J$.
The quantity $N_\gamma(E_{\rm x},E_{0})$ describes the number of photons at an energy $E_{\rm x}$ for a bremsstrahlung spectrum with endpoint energy $E_0$, and $\varepsilon_{abs}$ is the absolute efficiency at a given $E_{\rm x}$.
The product of both quantities is determined by normalizing a Monte-Carlo simulation of the bremstrahlung spectrum to well-determined $I_s$ values in $^{11}$B \cite{nndc}.

Two assumptions are made in order to convert the integrated cross sections to transition strengths: (i) the branching ratio $\Gamma_0/ \Gamma$ in Eq.~(\ref{eq:intc}) is put to one if no transitions to excited states are observed, i.e.\ possible unobserved decays to excited states are neglected, and (ii) all observed dipole transitions are assumed to be of $E1$ nature.
Assumption (i) has been shown to be on average a poor approximation at higher excitation energies \cite{ton10} but seems to hold reasonably well at excitation energies of $5 -7$ MeV \cite{isa13}.  
Although the investigated excitation region may have overlap with the spin-$M1$ resonance \cite{hey10}, approximation (ii) can be justified because even strong $M1$ transitions contribute little to the photoexcitation cross sections \cite{ton10,isa13}.  
The reduced transition probabilities can be extracted from the relation
\begin{equation}
    \label{eq:st}
   \frac{B(E1)\uparrow}{[{\rm e}^{2}{\rm fm}^{2}]}=9.554 \cdot 10^{-4} g \frac{\Gamma_{0}}{[{\rm meV}]} \cdot \left(\frac{[{\rm MeV}]}{E_{\rm x}}\right)^{3}.
\end{equation}

The resulting $B(E1)$ transition strengths for transitions in $^{112}$Sn and $^{120}$Sn are summarized in Tabs.~\ref{tab:tab1} and \ref{tab:tab2}, respectively. 
For each transition, the excitation energy $E_{\rm x}$, $\Gamma_{0}^2/\Gamma$, and the corresponding reduced transition probability are given.
The quoted uncertainties consider the statistical errors from the peak fit and systematic errors from the quantities entering into Eq.~(\ref{eq:pf}) except for an overall uncertantiy of the photon flux normalization not included, which is estimated to be about 10\%.   
 
%\begingroup 
%\squeezetable
%
%\pagebreak

%\begin{center}
\begin{longtable}{ccccccc}
%\begin{table}[tbh!]
\caption{Dipole transitions observed in $^{112}$Sn. \label{tab:tab1}} \\
%\begin{ruledtabular}
%\begin{tabular}{ccccccc}
\hline\hline
$E_{\rm x}$ & $\Gamma_{0}^{2}/\Gamma$ & B(E1)$\uparrow$ & & $E_{\rm x}$ &$\Gamma_{0}^{2}/\Gamma$ &B(E1)$\uparrow$\\
(keV) &(meV)&(10$^{-3}$e$^{2}$fm$^{2}$)&&(keV) &(meV) &(10$^{-3}$e$^{2}$fm$^{2}$) \\

\hline

3433.9   &162(15)     &11.5(11) & &6731.9   &289(51)     &2.7(5)\\

4141.3   &17(4)       &0.7(2)  &&6795.5   &185(25)     &1.7(2)\\

4162.3   &44(4)       &1.8(2)  &&6818.7   &139(23)     &1.3(2)\\

4330.4   &15(3)       &0.5(1)  &&6824.2   &194(32)     &1.7(3)\\

4726.5   &12(3)       &0.3(1)  &&6855.9   &170(25)     &1.5(2)\\

4837.4   &28(5)       &0.7(1)  &&6871.2   &189(19)     &1.7(2)\\

5057.1   &134(13)     &3.0(3)  &&6941.2   &367(41)     &3.1(3)\\

5128.2   &198(20)     &4.2(4)  &&6961.5   &362(53)     &3.1(5)\\

5246.2   &166(14)     &3.3(3)  &&6982.7   &246(30)     &2.1(3)\\

5480.5   &66(11)      &1.2(2)  &&7009.8   &62(15)      &0.5(1)\\

5502.6   &86(10)      &1.5(2)  &&7018.7   &82(16)      &0.7(1)\\

5593.7   &43(7)       &0.7(1)  &&7025.8   &86(17)      &0.7(1)\\

5617.6   &39(7)       &0.6(1)  &&7043.1   &245(42)     &2.0(3)\\

5649.1   &43(7)       &0.7(1)  &&7092.8   &524(48)     &4.2(4)\\

5666.4   &23(6)       &0.4(1)  &&7167.2   &363(42)     &2.8(3)\\

5699.9   &33(7)       &0.5(1)  &&7198.2   &578(75)     &4.4(6)\\

5748.6   &66(7)       &1.0(1)  &&7228.1   &164(27)     &1.2(2)\\

5812.7   &34(8)       &0.5(1)  &&7311.1   &138(28)     &1.0(2)\\

5860.7   &159(27)     &2.3(4)  &&7389.9   &183(30)     &1.3(2)\\

5884.0   &100(16)     &1.4(2)  &&7438.6   &275(42)     &1.9(3)\\

5924.1   &112(12)     &1.5(2)  &&7444.1   &233(37)     &1.6(3)\\

5976.6   &128(14)     &1.7(2)  &&7468.3   &186(45)     &1.3(3)\\

6005.0   &244(21)     &3.2(3)  &&7531.3   &429(62)     &2.9(4)\\

6059.8   &470(44)     &6.1(6)  &&7537.2   &770(82)     &5.2(6)\\

6080.9   &73(15)      &0.9(2)  &&7559.1   &323(43)     &2.1(3)\\

6096.9   &385(23)     &3.6(2)  &&7594.5   &205(31)     &1.3(2)\\

6129.0   &115(13)     &1.4(2)  &&7615.3   &257(41)     &1.7(3)\\

6150.4   &273(28)     &3.4(3)  &&7859.5   &207(35)     &1.2(2)\\

6168.3   &98(17)      &1.2(2)  &&7904.7   &196(40)     &1.1(2)\\

6198.7   &179(18)     &2.2(2)  &&7936.7   &272(39)     &1.6(2)\\

6224.3   &315(26)     &3.7(3)  &&7988.2   &606(62)     &3.4(3)\\

6246.4   &152(20)     &1.8(2)  &&8020.7   &412(67)     &2.3(4)\\

6259.1   &130(17)     &1.5(2)  &&8051.6   &396(60)     &2.2(3)\\

6272.6   &220(21)     &2.5(3)  &&8069.6   &482(65)     &2.6(4)\\

6313.3   &251(23)     &2.9(3)  &&8194.5   &518(75)     &2.7(4)\\

6348.7   &134(17)     &1.5(2)  &&8218.2   &262(48)     &1.4(2)\\

6388.1   &663(47)     &7.3(5)  &&8253.6   &177(38)     &0.9(2)\\

6404.1   &1686(120)   &18.4(13)&&8448.6   &147(41)     &0.7(2)\\

6428.6   &114(18)     &1.2(2)  &&8568.9   &166(43)     &0.8(2)\\

6450.0   &109(15)     &1.2(2)  &&8600.4   &118(35)     &0.5(2)\\

6520.7   &309(33)     &3.2(3)  &&8750.2   &249(56)     &1.1(2)\\

6550.1   &54(11)      &0.6(1)  &&8823.4   &278(64)     &1.2(3)\\

6601.0   &173(23)     &1.7(2)  &&9050.5   &413(108)    &1.6(4)\\

6679.9   &74(14)      &0.7(1)  &&9095.3   &268(65)     &1.0(2)\\

6706.7   &187(24)     &1.8(2)  &&9150.1   &240(75)     &0.9(3)\\

6715.0   &156(67)     &1.5(6)  &&9329.8   &599(138)    &2.1(5)\\
\hline\hline
%\end{tabular}
%\end{ruledtabular}
%\end{table}
\end{longtable}
%\end{center}

%\endgroup

%\squeezetable

\begin{longtable}{ccccccc}
%\begin{center}
%\begin{table}
\caption{Dipole transitions observed in $^{120}$Sn. \label{tab:tab2}} \\
%\begin{ruledtabular}
%\begin{tabular}{ccccccc}
\hline\hline
$E_{\rm x}$ & $\Gamma_{0}^{2}/\Gamma$ & B(E1)$\uparrow$ && $E_{\rm x}$ &$\Gamma_{0}^{2}/\Gamma$ &B(E1)$\uparrow$\\
(keV) &(meV)&(10$^{-3}$e$^{2}$fm$^{2}$)&&(keV) &(meV) &(10$^{-3}$e$^{2}$fm$^{2}$) \\
\hline

3279.4  &    137(14)  &8.6(9)  & &6432.3  &    142(28)  &1.5(3)\\

4251.0  &    73(10)   &2.7(4)  &&6443.7  &    299(52)  &3.2(6)\\

4564.8  &    36(8)    &1.0(2)  &&6469.7  &    375(62)  &4.0(7)\\

4679.7  &    52(10)   &1.5(3)  &&6485.8  &    409(67)  &4.3(7)\\

4939.0  &    36(8)    &0.9(2)  &&6520.7  &    186(32)  &1.9(3)\\

5245.4  &    22(7)    &0.4(1)  &&6539.5  &    219(40)  &2.2(4)\\

5354.4  &    37(13)   &0.7(2)  &&6644.3  &    438(68)  &4.3(7)\\

5408.2  &    54(13)   &1.0(2)  &&6691.0  &    206(41)  &2.0(4)\\

5447.2  &    126(21)  &2.2(4)  &&6727.3  &    238(55)  &2.2(5)\\

5638.0  &    109(18)  &1.8(3)  &&6898.9  &    508(163) &4.6(15)\\

5647.8  &    172(23)  &2.7(4)  &&6914.8  &    374(58)  &3.2(5)\\

5685.2  &    78(20)   &1.2(3)  &&6990.4  &    376(68)  &3.2(6)\\

5697.3  &    67(17)   &1.0(3)  &&7009.9  &    480(98)  &4.0(8)\\

5753.0  &    35(13)   &0.5(2)  &&7025.0  &    216(41)  &1.8(3)\\

5758.0  &    42(15)   &0.6(2)  &&7031.5  &    176(35)  &1.5(3)\\

5818.0  &    127(25)  &1.8(4)  &&7038.9  &    160(38)  &1.3(3)\\

5882.1  &    280(40)  &3.9(6)  &&7061.9  &    164(48)  &1.3(4)\\

5895.4  &    198(26)  &2.8(4)  &&7095.6  &    242(65)  &1.9(5)\\

5927.7  &    165(25)  &2.3(3)  &&7144.5  &    259(58)  &2.0(5)\\

5940.7  &    230(44)  &3.1(6)  &&7235.1  &    495(64)  &3.7(5)\\

5950.2  &    139(35)  &1.9(5)  &&7255.1  &    465(88)  &3.5(7)\\

5989.8  &    203(38)  &2.7(5)  &&7460.1  &    175(33)  &1.2(2)\\

6001.7  &    168(48)  &2.2(6)  &&7543.1  &    172(49)  &1.1(3)\\

6076.2  &    82(21)   &1.1(3)  &&7569.2  &    309(140) &2.0(9)\\

6093.5  &    110(24)  &1.4(3)  &&7624.9  &    190(40)  &1.2(3)\\

6127.1  &    248(35)  &3.1(4)  &&7701.2  &    229(57)  &1.4(4)\\

6152.5  &    127(23)  &1.6(3)  &&7889.0  &    312(62)  &1.8(4)\\

6252.4  &    255(48)  &3.0(6)  &&7958.6  &    523(93)  &3.0(5)\\

6267.0  &    350(44)  &4.1(5)  &&7975.6  &    606(98)  &3.4(6)\\

6285.8  &    160(31)  &1.8(3)  &&7994.5  &    237(48)  &1.3(3)\\

6305.9  &    270(37)  &3.1(4)  &&8044.3  &    120(30)  &0.7(2)\\

6332.6  &    363(54)  &4.1(6)  &&8079.7  &    258(100) &1.4(5)\\

6344.9  &    370(50)  &4.2(6)  &&8318.3  &    498(96)  &2.5(5)\\

6353.7  &    259(38)  &2.9(4)  &&8399.5  &    450(100) &2.2(5)\\

6375.0  &    118(23)  &1.3(3)  &&8478.3  &    304(80)  &1.4(4)\\

6397.0  &    240(40)  &2.6(4)  &&8554.9  &    447(139) &2.0(6)\\

6408.3  &    456(55)  &5.0(6)\\
\hline\hline
%\end{tabular}
%\end{ruledtabular}
%\end{table}
\end{longtable}
%\end{center}

\begin{figure}[tbh!]
    \includegraphics[angle=0,width=8.5cm]{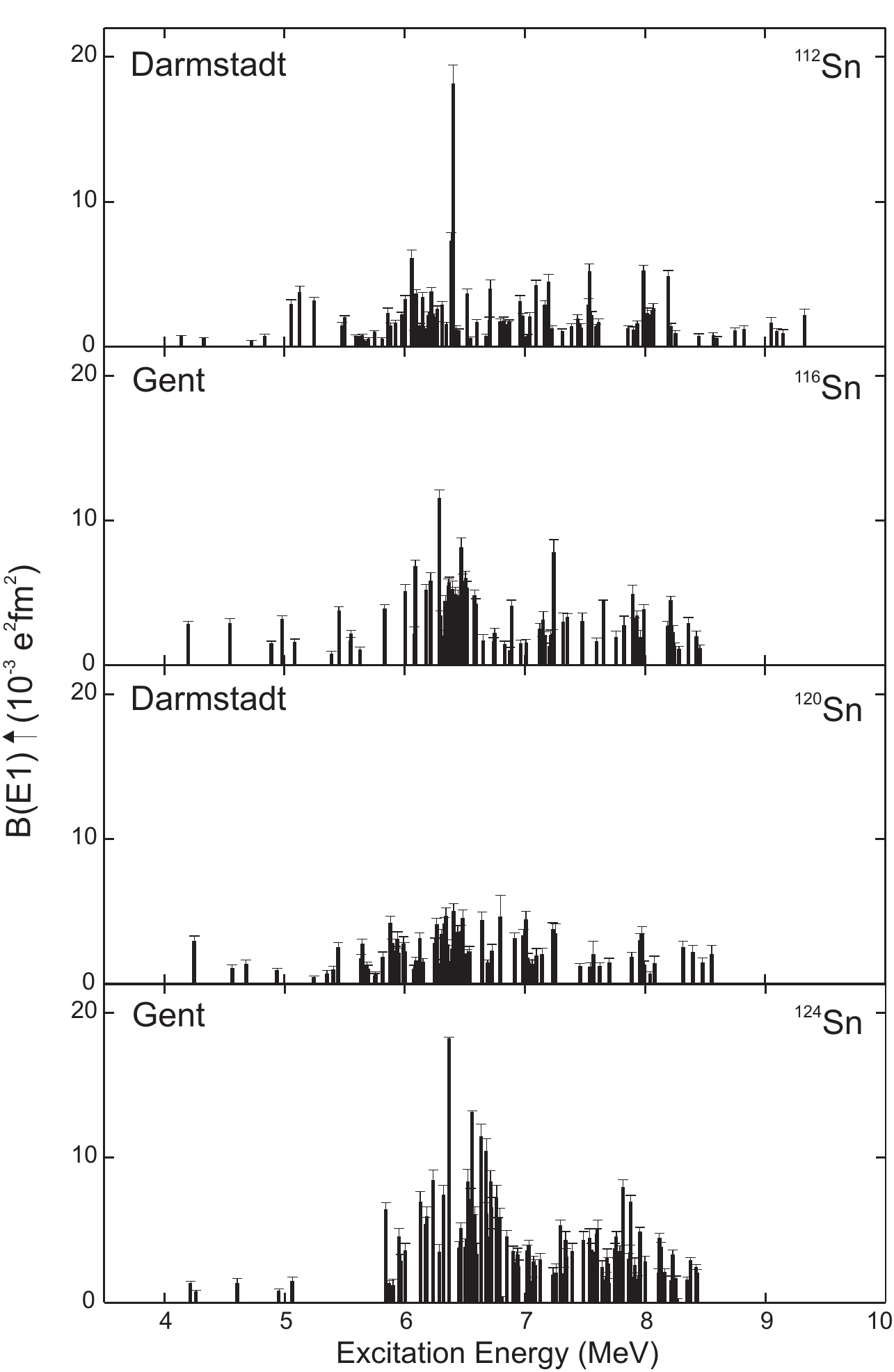}
        \caption{\label{fig:comp} Comparison of experimental $B(E1)$ strength distributions in $^{112,120}$Sn (present work) with $^{116,124}$Sn (from Ref.~\cite{gov98}).}
\end{figure}

In total 91 dipole transitions are observed for $^{112}$Sn up to 9.5 MeV endpoint energy with a summed  $B(E1$)$\uparrow$ transition strength of 0.187(25) e$^{2}$fm$^{2}$ corresponding to 0.25\% of the $E1$ energy-weighted sum rule (EWSR). 
The number of transitions in $^{120}$Sn is 72 and the summed $B(E1)$ transition strength amounts to 0.163(31) e$^{2}$fm$^{2}$ up to 9.1 MeV endpoint energy representing 0.22\% of the EWSR. 
Despite an experimental sensitivity limit up to 9 MeV comparable to the $^{112}$Sn measurement no transitions could be identified above 8.55 MeV.
The corresponding centroid energies of the low-energy $E1$ strength are 6.74 and 6.60 MeV for $^{112}$Sn and $^{120}$Sn, respectively. 
The two-phonon states \cite{bry99,pys06} are not included in the EWSR  and centroid values. 
A comparison with the previous NRF measurements on $^{116,124}$Sn \cite{gov98} up to 10 MeV endpoint is given in Tab.~\ref{tab:Sum}.
The centroid energy is roughly constant, but the total strengths found in $^{116,124}$Sn are larger than those from the present experiment.
Figure \ref{fig:comp}  presents a comparison of the $B(E1)$ strength distributions deduced for $^{112,120}$Sn in the present work with those of $^{116,124}$Sn from Ref.~\cite{gov98}.

%
%\begingroup
%\squeezetable
\begin{table}[tbh]
    \caption{Summed $B(E1)$ transition strengths and centroid energies of resolved transitions, and highest endpoint energy of the bremstrahlung spectra for $^{112,116,120,124}$Sn.}
\label{tab:Sum}
\begin{ruledtabular}
\begin{tabular}{cccc}
Isotope & $\sum B(E1) \! \uparrow$ (e$^{2}$fm$^{2}$) & $\bar{E}$ (MeV) & $E_0$ (MeV) \\
\hline
$^{112}$Sn   &0.175(24)&  6.7&  9.5\\
$^{116}$Sn   &0.233(28)&  6.7&  10.0\\
$^{120}$Sn   &0.164(31)&  6.6&  9.1\\
$^{124}$Sn   &0.379(45)&  7.0&  10.0\\
\end{tabular}
\end{ruledtabular}
\end{table}
%\endgroup
%

%

\subsection{Feeding}
\label{subsec:feeding}

\begin{figure}[tbh!]
    \includegraphics[angle=0,width=7cm]{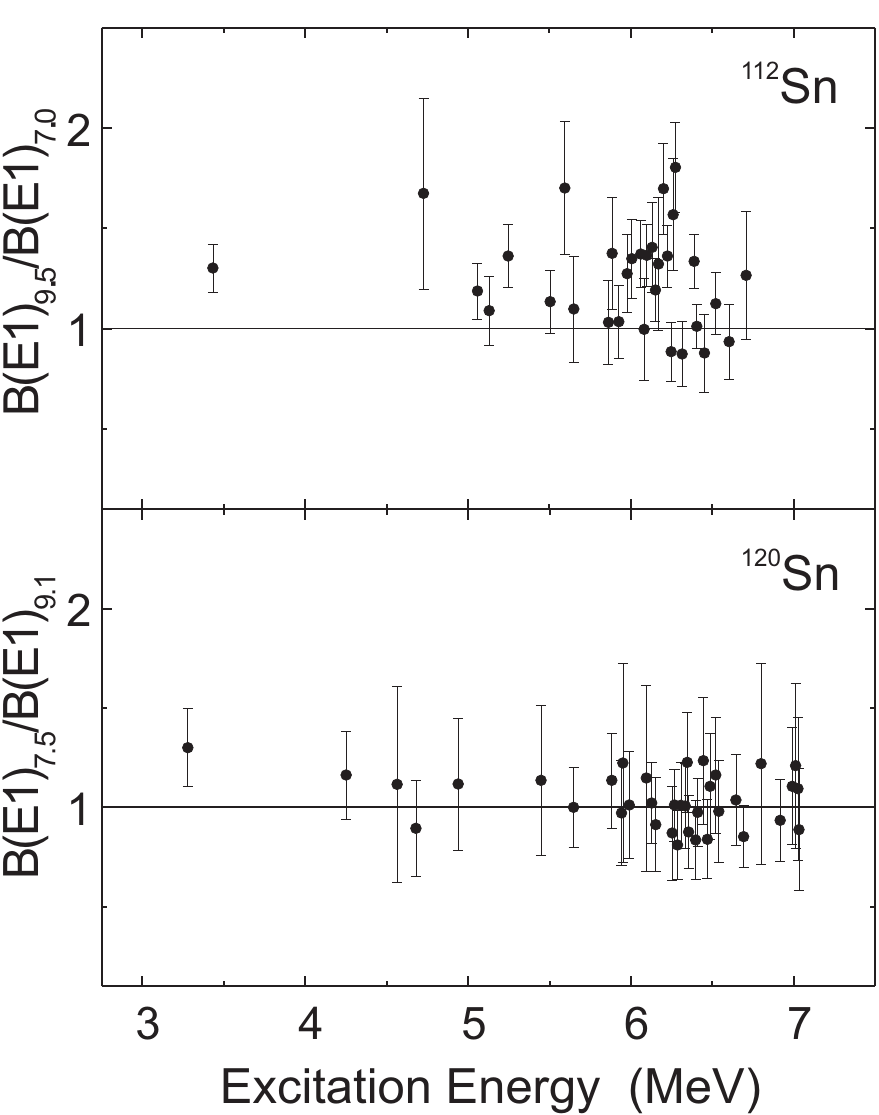}
        \caption{\label{fig:Ratio}Ratio of the $B(E1)$ transition strengths for $^{112}$Sn and $^{120}$Sn deduced at different endpoint energies.}
\end{figure}

The possibility of an indirect population of levels by feeding via inelastic transitions from higher-lying states needs to be corrected for the determination of g.s.\ transition strengths. 
This can be achieved by comparison of the reduced strength of a transition at different endpoint energies.
The ratio of $E1$ transition strengths at 9.5  and 7.0 MeV endpoint energies for $^{112}$Sn is displayed in the upper part of Fig.~\ref{fig:Ratio}. 
Note that the lower endpoint energy limits the excitation region of applicability and the number of transitions is smaller than in Tab.~\ref{tab:tab1}, because in some cases the signal in the spectrum taken at the lower endpoint energy was below the sensitivity limit.
Values larger than one in Fig.~\ref{fig:Ratio} within error bars indicate a feeding of the transition. 
This is clearly the case for transitions below 5 MeV, but also for a group around 6.5 MeV. 
For these cases, only the results obtained at 7.0 MeV endpoint energy enter into Tab.~\ref{tab:tab1}, while for the other transitions the results for both endpoint energies were averaged.
For $^{112}$Sn, also a spectrum measured at 5.0 MeV endpoint energy is available. 
The comparison between the results of 7.0  MeV and 5.0 MeV was discussed in Ref.~\cite{pol05}, and no feeding effect was observed.

The lower part of Fig.~\ref{fig:Ratio} shows the ratio of the transition strengths in $^{120}$Sn at 9.1 and 7.5 MeV endpoint energies, respectively.
The feeding pattern looks quite different from the case of $^{112}$Sn. 
Except for the lowest $E1$ transition populating the two-phonon $1^-$ state at 3.279 MeV, all ratios are consistent with one (i.e., no feeding)  within experimental uncertainties.
The results given in Tab.~\ref{tab:tab2} were averaged over both endpoint energies accordingly.

\subsection{Correction for branching ratios to excited states}
\label{subsec:BR}

As pointed out above, the results in Tabs.~\ref{tab:tab1} and \ref{tab:tab2} are derived under the assumption that branchings to excited states can be neglected. 
Thus, the $B(E1)$ strengths represent a lower limit only. 
Alternatively, one can try to correct for these branching ratios assuming statistical decay.
Such an approach is described e.g.\ in Ref.~\cite{rus08}, where photoexcitation and decay of the nucleus under investigation is modeled by average quantities ($\gamma$-ray strength functions and level densities). 
It is applied to discrete transitions only in the present case.

The simulations were performed for 100 nuclear realizations  of $^{112,120}$Sn using a fit of the $(\gamma,xn)$ data for the $E1$ strength function and the parametrizations from RIPL \cite{ripl} for the $M1$ and $E2$ strength functions.
Level densities were taken from back-shifted Fermi gas model (BSFG) fits. 
Two different parameterizations \cite{rau97,egi05} were tested to estimate the model dependence of the procedure.
The empirical approach of Rauscher {\it et al}. ~\cite{rau97} developed for $s$-process network calculations has been shown to be quite accurate in stable nuclei \cite{uts06,kal07}.
Von Egidy and Bucurescu \cite{egi05} developed a model, where the BSFG parameters are calculated from masses only.  

\begin{figure}[tbh!]
    \includegraphics[angle=0,width=8.5cm]{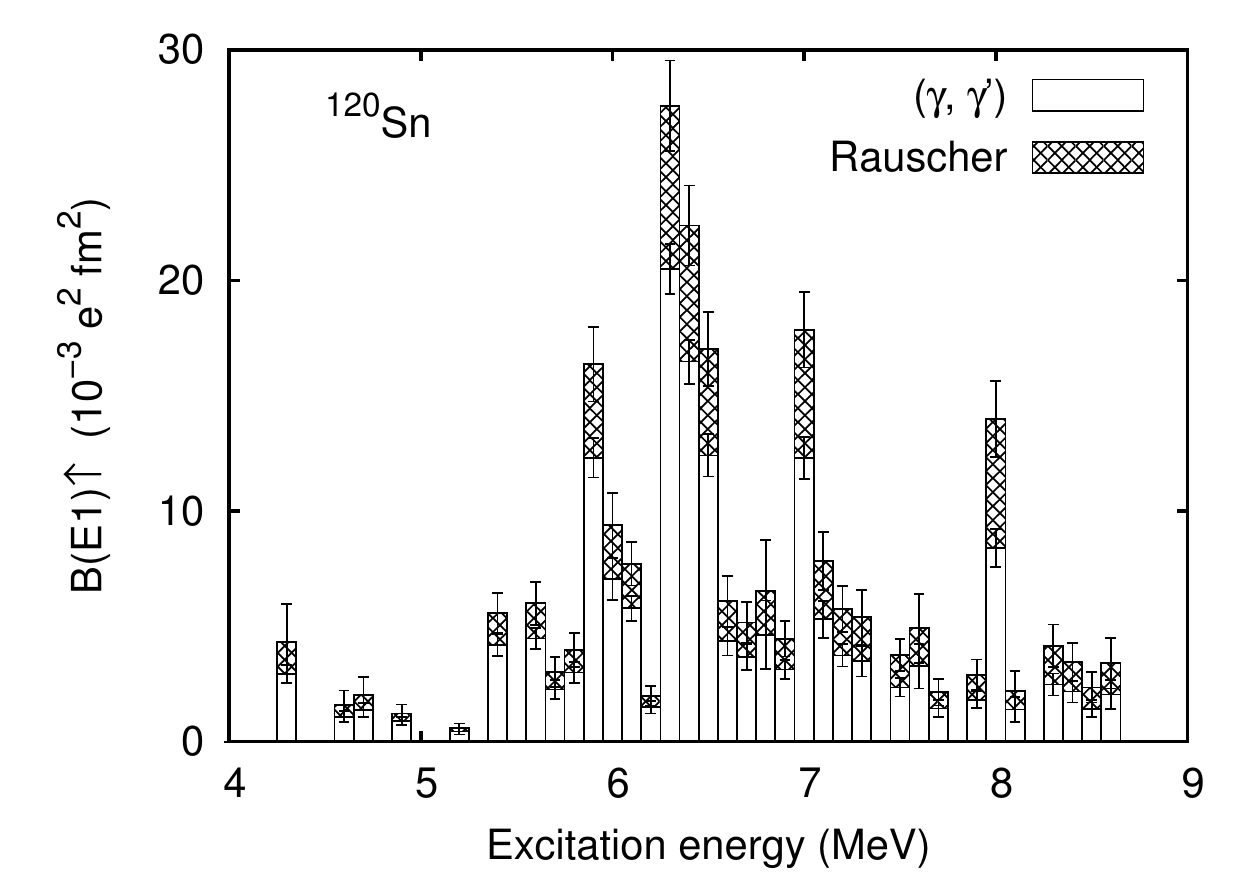}
        \caption{\label{fig:BRcorr}
$B(E1)$ strength distribution in $^{120}$Sn in 100 keV bins for the transitions given in Tab.~\ref{tab:tab2}. 
The open histogram gives the summed strengths, and the hatched areas indicate the statistical model correction using the level densities from Ref.~\cite{rau97}. }
\end{figure}

The resulting averaged branching ratios to excited states rise from about 10\% at $E_{\rm x} = 5$ MeV to about 40\% at $E_{\rm x} = 9$ MeV.
The predicted level densities in $^{120}$Sn, and correspondingly the branching ratios to excited states, are consistently higher than those in $^{112}$Sn. 
Differences between the two models arise from a steeper energy dependence but larger backshift parameter of Ref.~\cite{rau97} compared to Ref.~\cite{egi05}.  
Figure \ref{fig:BRcorr} shows, as an example, the B$(E1)$ strength distribution in $^{120}$Sn in 100 keV bins with and without inclusion of the branching ratio correction.

\begin{figure}[tbh!]
    \includegraphics[angle=0,width=7cm]{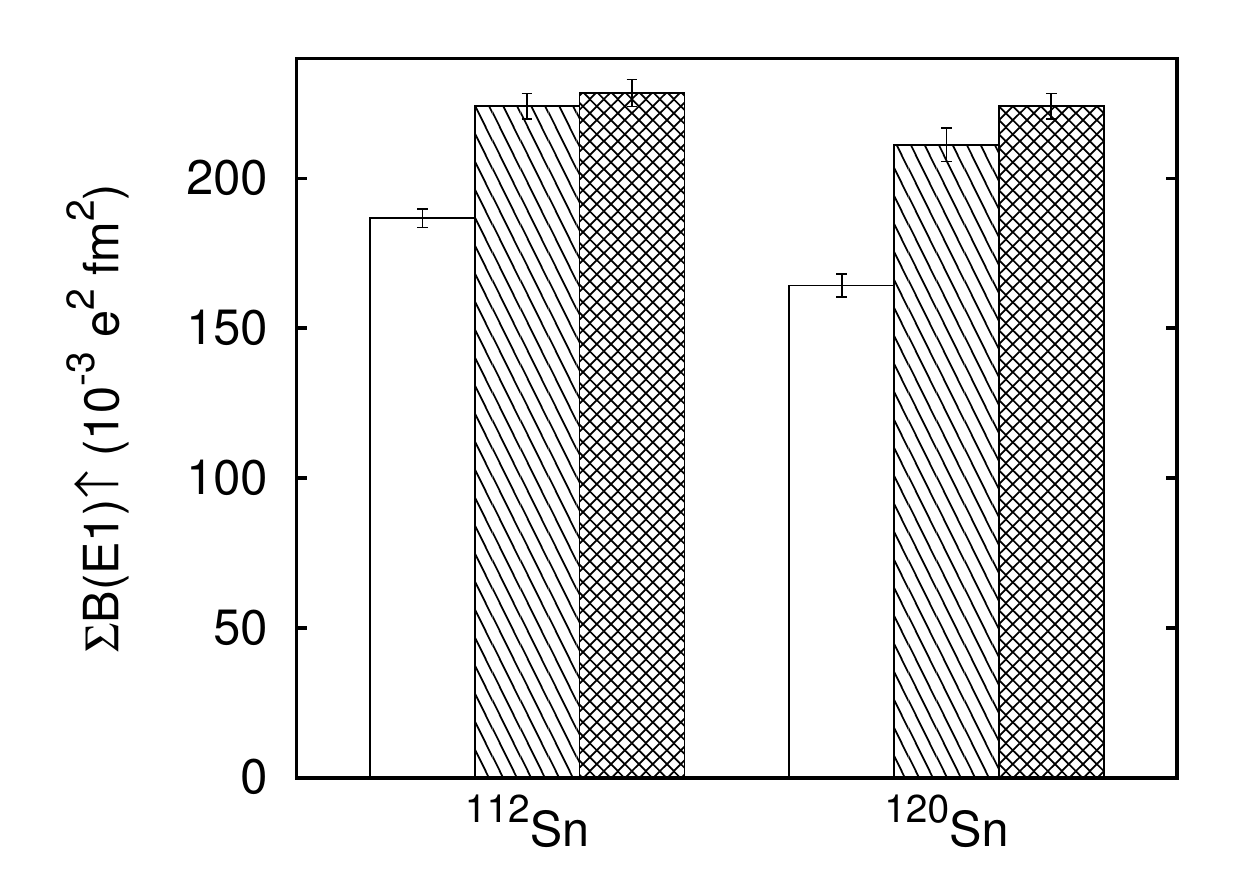}
        \caption{\label{fig:E1corr} 
Summed B$(E1)$ strengths in $^{112,120}$Sn from the present experiment without (open bars) and with correction for branching ratios using the model of Ref.~\cite{rus08} and level density parameters of Ref.~\cite{egi05} (light hatched bars) and Ref.~\cite{rau97} (dark hatched bars).}
\end{figure}
The impact of the correction on the summed $E1$ strengths of Tab.~\ref{tab:Sum} is illustrated in Fig.~\ref{fig:E1corr}.
The dependence on the chosen level density model is small for $^{112}$Sn and somewhat larger for $^{120}$Sn. 
For $^{112}$Sn an increase of 23\%(20\%) and for $^{120}$Sn of 39\%(29\%) is observed with the parameters of Ref.~\cite{rau97}(Ref.~\cite{egi05}).
Even including the correction the summed strengths are still smaller than those in $^{116,124}$Sn \cite{gov98} without the correction. 
This finding is not influenced by the slightly higher endpoint energies of the experiments in Ref.~\cite{gov98}, since the largest excitation energy of analyzed transitions is comparable in both experiments.
However, it should be emphazised that the simulated average branching ratios can only serve as a guidance since the low-energy part of the photon strength function entering the analysis might play an important role \cite{isa13}.
In order to further investigate this important question, $(\gamma,\gamma' \gamma'')$ coincidence experiments with a new setup \cite{loh13} are underway.

\section{Fluctuation Analysis}
\label{sec:fluc}

According to theoretical predictions discussed in the Introduction, the summed B($E1$) strengths of the even-mass tin isotopes in the low-energy region should increase with the number of neutrons. 
However, the experimental results show that  $^{120}$Sn has the lowest summed strength, and also its strength distributions is more fragmented compared to the other even-mass stable tin isotopes. 
One possible explanation would be unresolved strength hidden in the background. 
To determine unresolved strength, a fluctuation analysis was applied to ($\gamma,\gamma^{\prime}$) spectra in order to investigate the $M1$ scissors mode strength in deformed odd-mass nuclei \cite{hey10,end97,end98,hux99,nor03}. 
In this section the method is explained and applied to the ($\gamma, \gamma^{\prime}$) spectra of $^{112,120}$Sn.

\subsection{Method}
\label{subsec:fluc-method}

\begin{figure}[tbh!]
\includegraphics[angle=0,width=8.5cm]{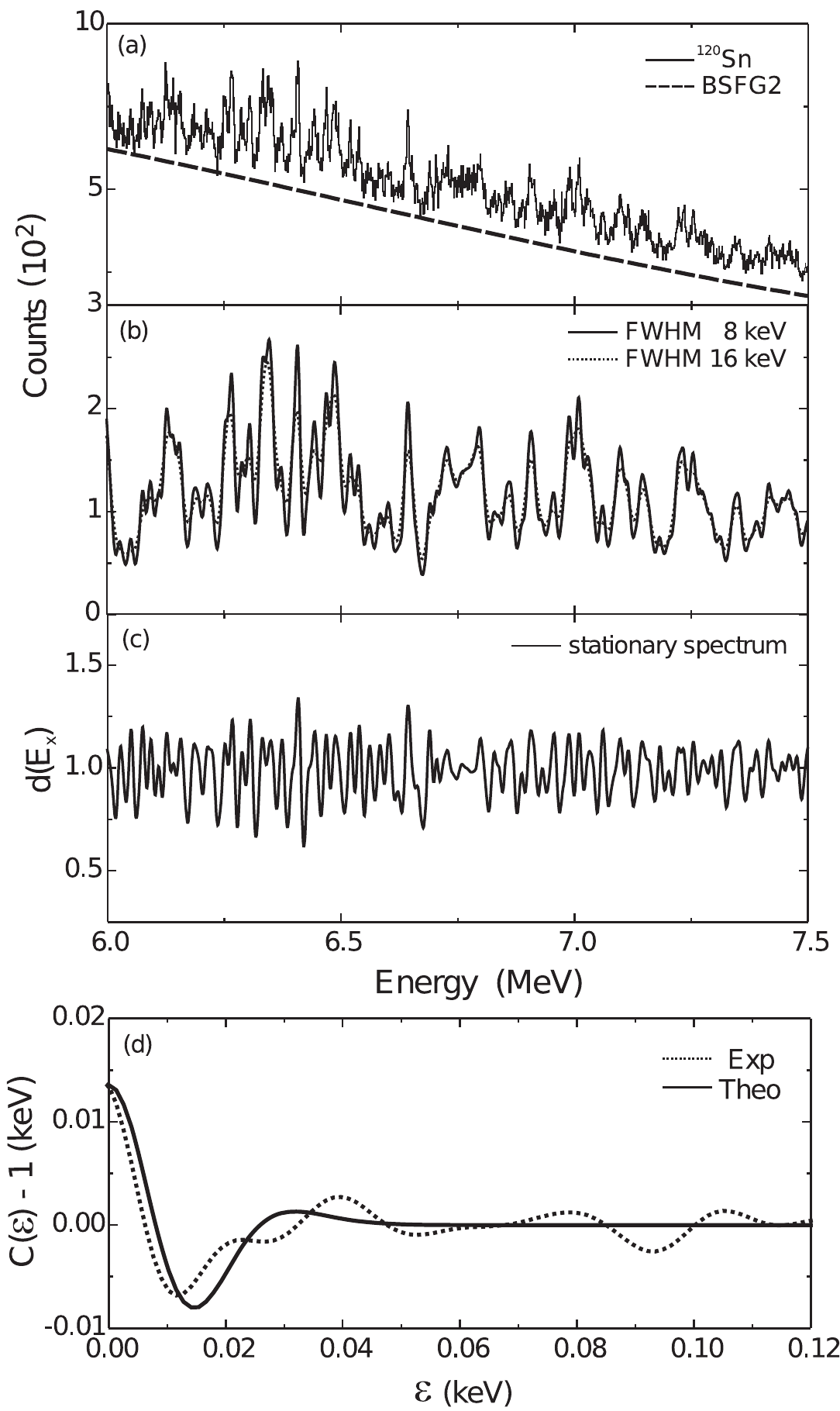}
\caption{\label{fig:fluc}The steps of the fluctuation analysis: (a) Spectrum of the $^{120}$Sn($\gamma,\gamma^{\prime}$) reaction and selected background, (b) smoothed spectra with 16 keV and 50 keV FWHM, (c) stationary spectrum $\textit{d($E_{\rm x}$)}$, (d) theoretical and experimental autocorrelation functions as a function of the shift parameter $\epsilon$.}
\end{figure}

The method is applicable in the excitation energy region where the mean level spacing $\langle D \rangle$ is smaller than the experimental energy resolution $\Delta E$ and at the same time the mean level width $\langle \Gamma \rangle$ is smaller than $\langle D \rangle$ and $\Delta E$
\begin{equation}
    \label{eq:DelE}
    \langle \Gamma \rangle \leq \langle D \rangle < \Delta E.
\end{equation}
The fluctuations in the measured spectra should also be related to the ground-state transition widths $\textit{$\Gamma_{0}$}$ only. 
Thus one has to remove the transitions from the $^{11}$B calibration standard as well as all single- and double-escape peaks.
This is achieved by subtracting the respective peak from a smooth background whose energy dependence is determined from a local fit.
Since no inelastic transitions were observed, it is reasonable to assume that peaks resulting from the branching to excited states are small enough not to contribute to the fluctuations.
For typical branching ratios predicted within the statistical model described in Sec.~\ref{subsec:BR} this condition is fulfilled.
Then one can extract the  unresolved strength from the fluctuations of a spectrum applying the steps described in Fig.~\ref{fig:fluc}. 

Panel \ref{fig:fluc}(a) shows the spectrum with a backgorund (dashed line) determined as described below.
The background subtracted spectrum $g_>(x)$ is smoothed by convolution with a Gaussian function with width $\sigma_{>}$ to remove gross structure . 
The optimum value of $\sigma_{>}$ is chosen such that small variations around this value do not change the results.
The resulting spectrum is shown as dashed line in Fig.~\ref{fig:fluc}(b).
In order to diminish the contribution of counting statistics to the fluctuations, the spectrum is also folded with a Gaussian function with width $\sigma_{<}$ smaller than the experimental energy resolution which produces the $\textit{$g_{<}(x)$}$ spectrum shown in Fig.~\ref{fig:fluc}(b) by the solid line. 
The so-called stationary spectrum $\textit{d($E_{\rm x}$)}$ is defined as the ratio of the $\textit{$g_{>}(x)$}$ and $\textit{$g_{<}(x)$}$ spectra and shows local fluctuations in a given energy interval displayed in Fig.~\ref{fig:fluc}(c).

A measure of the fluctuations is given by the autocorrelation function of the stationary spectrum
\begin{equation}
\label{eq:autoco-exp}
C(\epsilon) = \langle d(E_{\rm x})d(E_{\rm x}+ \epsilon) \rangle,
\end{equation}
where the brackets indicate averaging over the interval for which the analysis is performed, and $\textit{$\epsilon$}$ is the shift parameter in the autocorrelation. 
The experimental autocorrelation function can be well approximated \cite{jon76,han79,han90} by the analytical expression
\begin{equation}
\label{eq:autoco-th}
C(\epsilon)-1 = \frac{\alpha \cdot \langle D \rangle }{2\sigma
\alpha \sqrt{\pi}}\times f(\varepsilon, \sigma_{>},\sigma_{<}),
% \{\exp(\frac{-\epsilon^{2}}{4\sigma_{<}^{2}})+
%\frac{1}{y}\cdot \exp{- \frac{\epsilon^{2}}{4
%\sigma_{<}^{2}y^{2}}} -\sqrt{\frac{8}{1+y^{2}}}\cdot \exp(-
%\frac{\epsilon^{2}}{4 \sigma_{<}^{2}(1+y^{2})}) \}
\end{equation}
where $\langle D \rangle$ is the average level spacing and the function $f$ depends on experimental parameters only. 
The background is now varied until the theoretical and experimental autocorrelations agree at $\textit{$\epsilon$}$ = 0, where there is a simple linear relation between $\textit{C(0)}$ and the product $\alpha$$\langle$D$\rangle$ since the function $\textit{f}$ is normalized $\textit{f}$($\epsilon=0$)=1. 
The variance $\alpha$ depends on the statistical distributions of both the level spacings and intensities.
For $E1$ transitions and excitation energies below 7 MeV in heavy nuclei evidence was found that these are close to the Wigner and Porter-Thomas distributions, respectively \cite{end04}.

The experimental and theoretical autocorrelation functions are plotted in Fig.~\ref{fig:fluc} (d). 
Differences between the  experimental and theoretical autocorrelation functions for finite $\epsilon$ may be due to errors induced by the finite range of the energy interval.

\subsection{Background determination}

The relation between the value of the autocorrelation function at $\epsilon = 0$ and the average level spacing $\langle D \rangle$ established by Eq.~(\ref{eq:autoco-th}) can be utilized in two ways.
Provided the background in the spectra can be estimated, the fluctuation analysis allows to extract $\langle D \rangle$ and thus the level density $\rho = 1/\langle D \rangle$.
This method has been applied successfully to extract level densities from the fine structure of giant resonances \cite{kal06,kal07,usm11,pol14}.
The background was determined from a wavelet decomposition of the spectra using discrete wavelet transforms \cite{she08}.
However, the method is not applicable in the present case because it requires a compact resonance and a good peak-to-background ratio, while the $(\gamma,\gamma')$ spectra show highly fragmented strength on top of a very large background from atomic processes.

Alternatively, if the level density is known experimentally or estimated by a model, one can determine the amount of background needed such that the value $C(0)$ from Eq.~(\ref{eq:autoco-th}) matches the experimental result from Eq.~(\ref{eq:autoco-exp}).
In the present case we rely on the empirical parameterizations of Refs.~\cite{rau97} and \cite{egi05} within the backshifted Fermi-gas model called BSFG1 and BSFG2, respectively, hereafter.
These models have also been used to estimate the branching ratio of unobserved decays to states other than the ground state in Sec.~\ref{subsec:BR}.
Additionally, a microscopic statistical model is used to obtain the nuclear level densities. 
It is based on the ground-state structure properties predicted within the Hartree-Fock-BCS (HF-BCS) approach \cite{dem01} and includes a consistent treatment of the shell effects, pairing correlations, deformation and collective excitations. 
The variation of the level-density input will permit to estimate the model dependence of the background determination.

\subsection{Application to Photon Scattering Spectra}

\begin{figure}[tbh!]
    \includegraphics[angle=0,width=8.5cm]{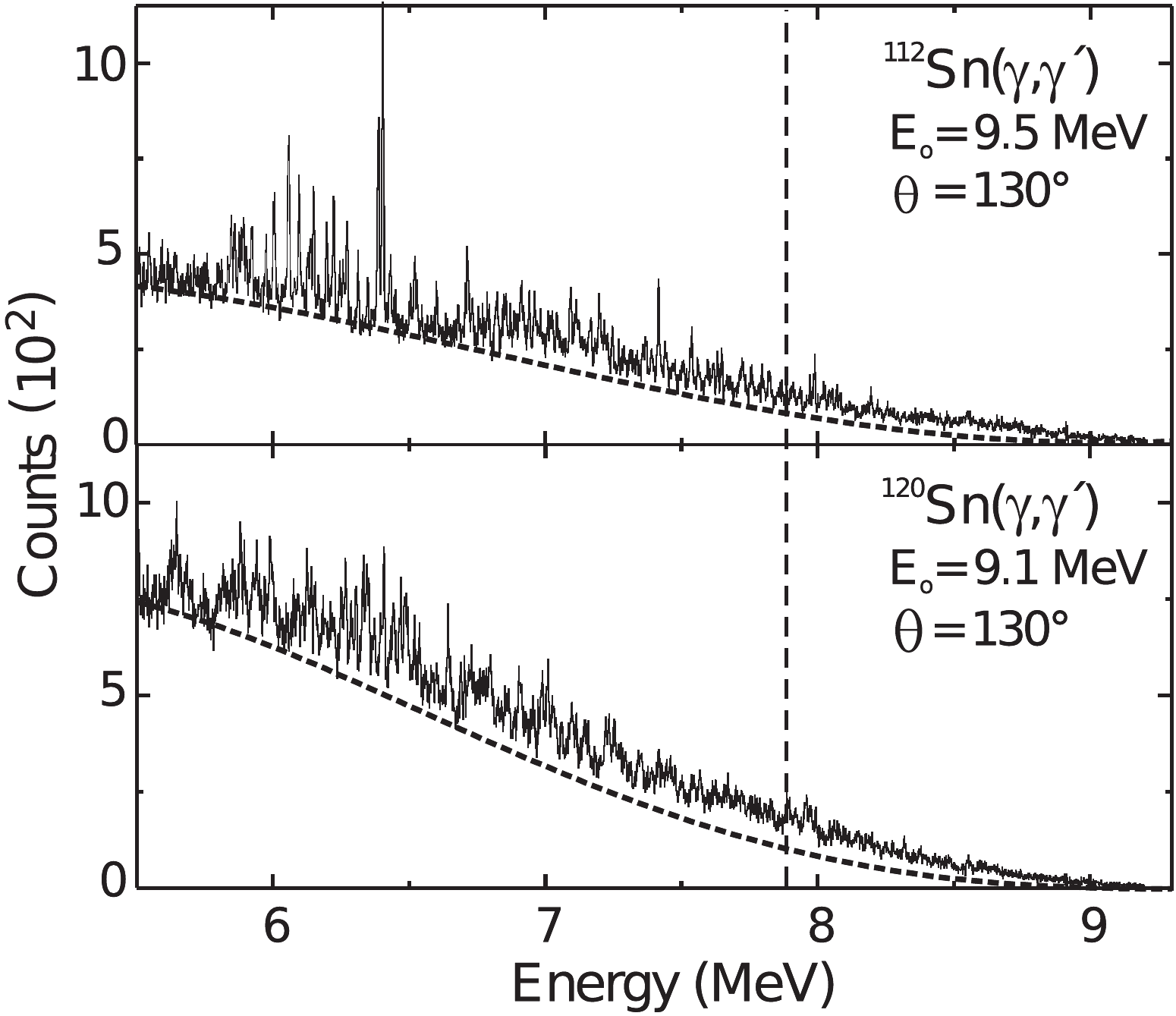}
%    \label{fig:Bgr}
        \caption{\label{fig:Bgr}
$^{112}$Sn and $^{120}$Sn spectra at 130$^{\circ}$ and backgrounds deduced from the fluctuation analysis described in the text for the BSFG1 \cite{rau97}, BSFG2 \cite{egi05} and HF-BCS \cite{dem01} level density models.
Note that the resulting background curves are indistinguishable within the line size.}
\end{figure}

For the inverse application of the fluctuation analysis described above, the excitation energy region is divided in intervals of 200 keV, which are small enough to assume a constant background. 
These intervals are individually fitted and then a smooth curve using a spline function is drawn through points of the background count rates defined by the center of the intervals. 
An example using the BSFG2 model is shown in Fig.~\ref{fig:fluc}(a). 
This process is applied to the spectra at 90$^{\circ}$ and 130$^{\circ}$ and for both isotopes.
The curves resulting  for the three level density models are shown in Fig.~\ref{fig:Bgr} by way of example for the $130^\circ$ spectra.  
Although the models predict different absolut values and energy dependences of the level densities, the resulting background curves are indistinguishable in Fig.~\ref{fig:fluc}(a).
Obvioulsy the value of the autocorrelation function at $\epsilon = 0$ is much more sensitive to the area under the background than to its shape.

There are constraints to the method which limit the applicability in the present spectra to a region $E_{\rm x} =5.5 - 7.8$ MeV. 
At lower excitation energies, condition (\ref{eq:DelE}) is not fulfilled.
Although the background curves in Fig.~\ref{fig:Bgr} extend to higher energies, for $E_{\rm x} > 7.8$ MeV the signal gets too weak because of a lack of statistics and/or because the average widths start to overlap. 

To determine the strength, the spectrum is unfolded with help of a GEANT4 simulation \cite{has07}  to account for the Ge detector response.
Starting from the highest energy, the contributions at lower energies due to single-escape events and Compton scattering are subtracted. 
The remaining area above the background is integrated and converted to a $B(E1)$ transition strength, which includes the contributions from resolved (Sec.~\ref{subsec:BE1}) and unresolved transitions.  
The procedure is repeated for each level density model and the two measured scattering angles.
The resulting strengths at 90$^{\circ}$ and 130$^{\circ}$ are consistent within the respective error bars for both isotopes demonstrating the reliability of the analysis method. 
The final result is obtained by averaging over the results from the two angles for each level density model and finally averaging over results for the different level density models.

The resulting total strengths  and the decomposition into unresolved and resolved contributions (based on the analysis of Sec.~\ref{subsec:BE1}) are summarized in Tab.~\ref{tab:flucres}. 
The errors of the unresolved and total strengths include uncertainties due to variation of the parameters of the fluctuation analysis (see, e.g., Ref.~\cite{usm11}) and due to the differences between the results from the three level density models. 
\begingroup
%\squeezetable
\begin{table}[t]
    \caption{Total $B(E1)$ strengths (in e$^{2}$fm$^{2}$) deduced from the $^{112,120}$Sn$(\gamma,\gamma')$ experiments and contributions from discrete peaks and from the fluctuation analysis up to $E_{\rm x} = 7.8$ MeV.}
    \label{tab:flucres}
\begin{ruledtabular}
\begin{tabular}{rccc}
$\sum B(E1)$ & total  & fluctuation & discrete \\
            \hline
$^{112}$Sn   &0.255(16)&    0.113(16)  &0.142(18)\\
$^{120}$Sn   &0.253(33)&    0.120(33)    &0.133(25)
 \end{tabular}
        \end{ruledtabular}
\end{table}
\endgroup

The unresolved strength amounts to 44 $\%$ and 47 $\%$ of the total $B(E1)$ strength up to $E_{\rm x} = 7.8$ MeV extracted from the $(\gamma,\gamma')$ spectra in  $^{112}$Sn and $^{120}$Sn , respectively. 
This result can be affected by a possible difference between average branching ratios for strong (resolved strength) and weak (unresolved strength) tranisitons (see also Sec.~\ref{subsec:BR}).
We reiterate that the unresolved strength could only be extracted in the energy region between 5.5 and 7.8 MeV and thus represents a lower limit only when compared to the total strength due to discrete transitions given in Tab.~\ref{tab:Sum}. 
The absolute amount of unresolved strength is very similar in both isotopes.
The differences between the three level density predictions in the investigated excitation energy region ($<$ 20\% for $^{112}$Sn and 50 - 100\% for $^{120}$Sn) seem to have a minor impact. 

\section{Comparison with Theoretical Models}
\label{sec:comp}

In this section, the measured $B(E1$) transition strengths below the neutron threshold in $^{112,120}$Sn are compared to theoretical appraoches. 
As discussed in the Introduction, predictions of the low-energy $E1$ strength are available from a variety of mean-field models.
However, a realistic description of the strongly fragmented experimental strength distributions requires the inclusion of complex configurations.
Therefore, the comparison focuses on two approaches, QPM and RQTBA, which allow  to go beyond the $1p1h$ level and include $2p2h$ or even $3p3h$ states.

Basics of the QPM are described in Ref.~\cite{sol00}.
In the current work, two QPM calculations are presented called QPM Darmstadt and QPM Giessen hereafter.
While both calculations include the coupling of 2- and 3-phonon states in a similar way, they differ in the way how the underlying mean field and parameters of the residual interaction are determined.
In the case of QPM Darmstadt single-particle energies stem from a global Woods-Saxon parameterization obtained from a fit to experimental data over a wide mass range \cite{pon79}.
These are further modified (typically up to a few hundred keV) by adjusting to experimental values obtained from the odd-mass neighboring nuclei.
Strength parameters of the residual interaction are fixed by the properties of the lowest collective vibrations in the respective nucleus.
This approach has been shown to provide a very good description of collective phenomena in vibrational nuclei (see e.g.\ Refs.~\cite{rye02,pon99,vnc99,she04,sav08,wal11}).     
The QPM Giessen calculation is based on a selfconsistent derivation of the mean-field properties in a Hartee-Fock-Bogoliubov approach (although with some empirical adjustments) described in Ref.~\cite{tso08} and has been shown to provide a good description of low-energy dipole strength in semimagic nuclei \cite{ton10,sch13,rus13}.  
Both calculations include the full 2- and 3-phonon space resulting from the coupling of $J^\pi = 1^\pm - 7^\pm$ phonons up to $E_x = 9$ MeV.  

The RQTBA results are based on a relativistic mean-field approach and allow a selfconsistent calculation of the $E1$ response.
Two ways of including configurations beyond the $1p1h$ level have been presented recently.
In Ref.~\cite{lit07} an extension of the approach to include 2-quasiparticle$\otimes$phonon states (called 2qp+phonon RQTBA) is described and an application to the $E1$ strength in the tin isotope chain is discussed in Ref.~\cite{lit08}.
Recently, the model has been extended to include the 2-phonon model space (called 2-phonon RQTBA) resulting from the coupling of natural-parity phonons up to $J = 6$ and shown to impact on the description of the low-energy $E1$ strength in stable tin isotopes \cite{lit10,lit13}.   

\subsection{$E1$ strength distributions}

\begin{figure}[tbh!]
    \includegraphics[angle=0,width=7cm]{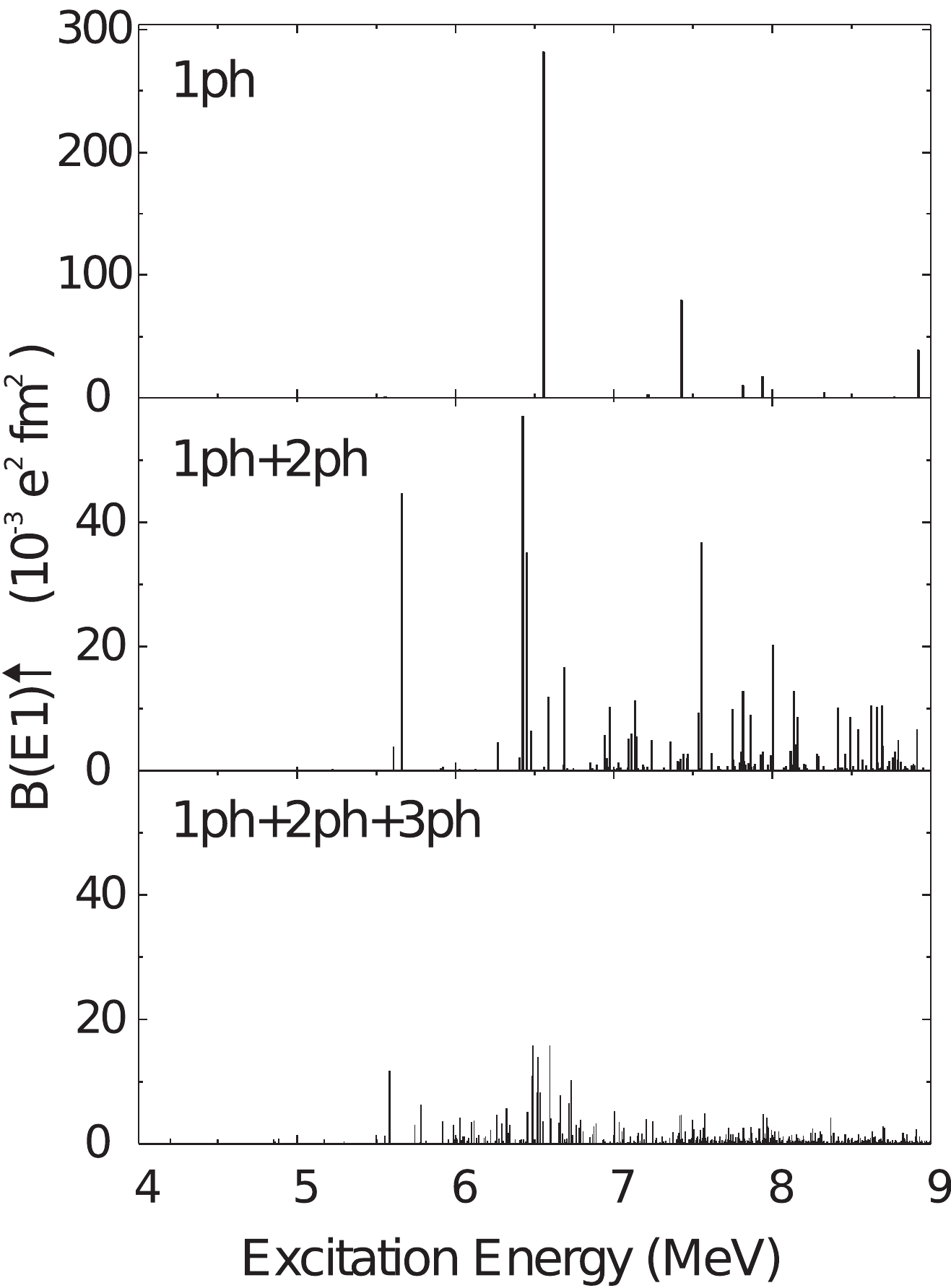}
        \caption{\label{fig:1p2p3p}Low-energy $B(E1)$ strength in $^{120}$Sn predicted with the QPM Darmstadt approach for 1-phonon, (1+2)-phonon, and (1+2+3)-phonon model spaces.
Note the differences in the absolute scales.}
\end{figure}
\begin{figure*}[tbh]
%\begin{figure}[tbh!]
    \includegraphics[angle=0,width=16cm]{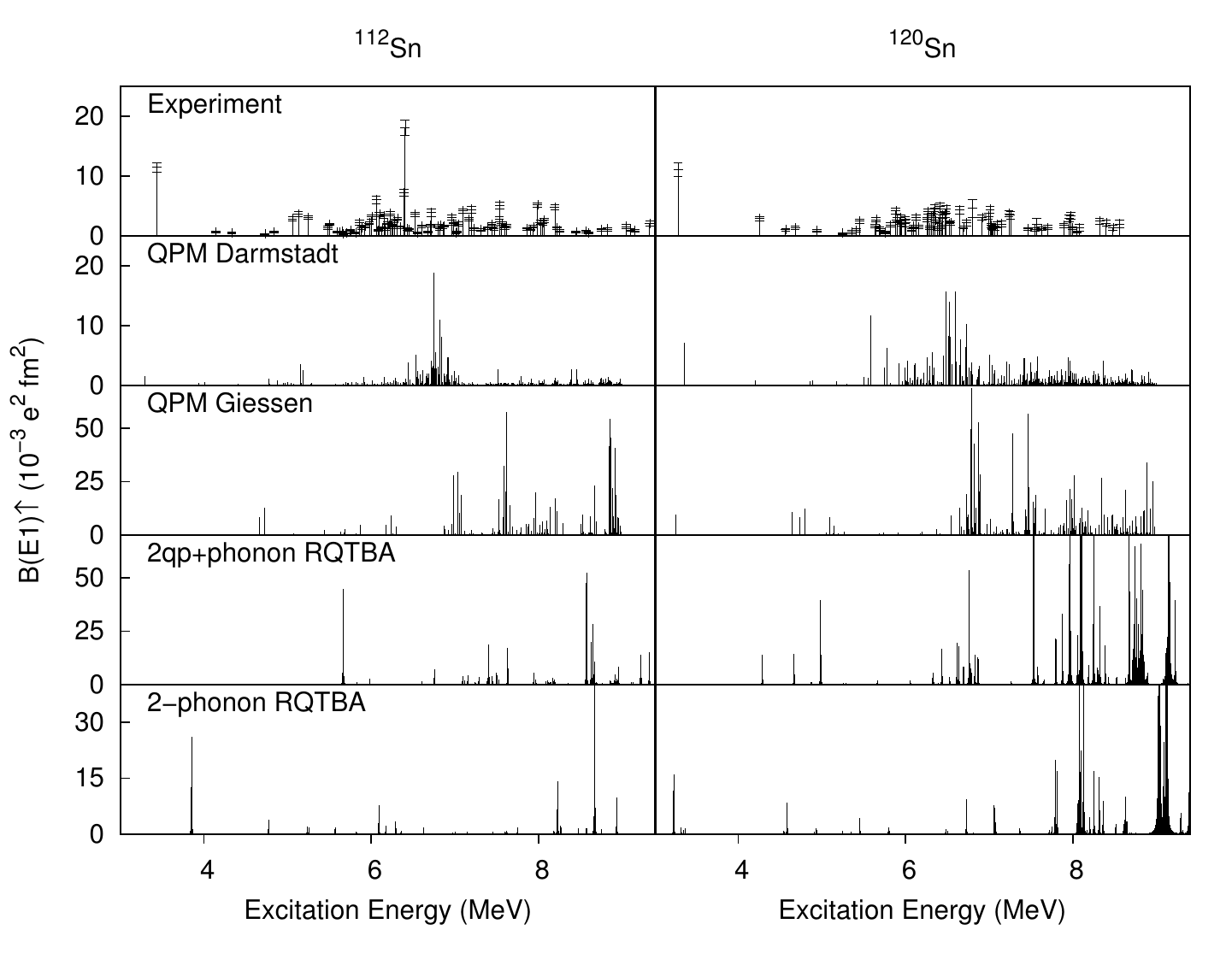}
        \caption{\label{fig:theory} Experimental  $B(E1)\! \uparrow$ strength distributions in $^{112}$Sn (left) and $^{120}$Sn (right) up to $E_x = 9$ MeV in comparison with model calculations described in the text.
Note that the differences of absolute scales.} 
%\end{figure}
\end{figure*}

Before comparing the different models with experiment, we illustrate in Fig.~\ref{fig:1p2p3p} the effect of including complex configurations beyond $1p1h$ by a calculation of the $B(E1)$ strength distribution in $^{120}$Sn (using by way of example QPM Darmstadt) allowing for 1-phonon, (1+2)-phonon, and (1+2+3)-phonon model spaces, respectively. 
At the 1-phonon level, the strength distribution below 9 MeV is dominated by two transitions only.
When going to a (1+2)-phonon model space, considerable fragmentation is observed (note the scale change in Fig.~\ref{fig:1p2p3p}).
The strength of the most prominent transition is reduced by about a factor of five and some strength is shifted to lower energies, while the total strength remains approximately the same.
In the full (1+2+3)-phonon calculation the average transition strength is reduced by another factor four to five, again without changing the total $B(E1)$ strength.
As discussed below, only with the inclusion of 3-phonon states a realistic quantitative reproduction of the fragmentation can be achieved.     

The measured (discrete transitions only) and predicted $B(E1)$ strength distributions for $^{112}$Sn and $^{120}$Sn are compared in Fig.~\ref{fig:theory}.
The QPM Darmstadt calculations provide a satisfactory agreement of the fine structure for both cases.
In $^{120}$Sn the strongest model transitions show about a factor of two larger $B(E1)$ values than seen in the data.
However, experimentally a stronger fragmentation is observed in $^{120}$Sn compared to $^{112,116,124}$Sn  (cf.\ Fig.~\ref{fig:comp}).
In the QPM Giessen calculations a prominent cluster of transitions with a strength comparable to experiment is observed for both nuclei around the experimental centroid energy. 
At higher excitation energies the predicted strength is large compared to the data.
In general, the strength is less fragmented than in the QPM Darmstadt calculation despite comparable model spaces.

The RQTBA results differ substantially from each other.
The 2qp+phonon version predicts more strength below 6 MeV than seen in the experiments.
For $^{120}$Sn, a bump roughly at the experimental centroid energy is observed but overestimates the strength.
The strength at higher $E_x$ is much larger than in all other calculations.   
The 2-phonon RQTBA shows less fragmentation  and a shift of the very strong transitions found in the 2qp+phonon calculation above 7 MeV to higher excitation energies.
These findings can be related to the geometrical properties of the phonon amplitudes (cf.\ Eq.~(C4) in Ref.~\cite{lit08} and Eqs.~(28,29) in Ref.~\cite{lit13}, respectively).  
Overall, this clearly improves the comparison to experiment \cite{lit10,lit13}.

%
%\begingroup
%\squeezetable
\begin{table}[tbh!]
    \caption{$B(E1)$ transition strengths (in e$^2$fm$^2$) for $^{112,120}$Sn summed over excitation energy regions $4-8$ MeV and $4 -9$ MeV.} \label{tab:Sumtheo}
\begin{ruledtabular}
\begin{tabular}{lcccc}
& \multicolumn{2}{c}{$^{112}$Sn} & \multicolumn{2}{c}{$^{120}$Sn} \\
& 4-8 MeV & 4-9 MeV & 4-8 MeV & 4-9 MeV \\
\hline
Experiment & 0.163 & 0.181 & 0.154 & 0.164 \\
QPM Darmstadt & 0.213 & 0.374 & 0.399 & 0.553 \\
QPM Giessen & 0.445 & 0.933 & 0.887 & 1.364 \\
2qp+phonon RQTBA & 0.622 & 1.511 & 3.908 & 9.494 \\ 
2-phonon RQTBA & 0.226 & 0.743 & 0.583 & 2.345  
 \end{tabular}
        \end{ruledtabular}
\end{table}
%\endgroup
%

Table \ref{tab:Sumtheo} collects the summed $B(E1)$ strengths from experiment and the various models.
In general, the model strengths integrated over the experimentally accessible excitation region up to 9 MeV are much larger than measured in this work.
However, one should recognize that the experimental numbers do not include the unresolved part which almost doubles the $B(E1)$ values (cf.\ Tab.~\ref{tab:flucres}).
Furthermore, the fluctuation analysis described in the previous section was limited to excitation energies up to 7.8 MeV, and it can be expected that the contributions at higher $E_x$ are even larger.
Also, the branching ratios to excited states are neglected although the corrections are expected to be of the order $20 -40$\% only in these semimagic nuclei (cf.\ Sec.~\ref{subsec:BR} ).
The comparison also shows strong sensitivity to the upper excitation energy limit.
For example, restricting the upper limit of summation to 8 MeV, QPM Darmstadt and 2-phonon RQTBA describe the experimental strength in $^{112}$Sn quite well. 
For $^{120}$Sn, these models are somewhat too high but closer to the data than the other predictions.  

The differences found between the model approaches (in particular, between QPM Darmstadt and QPM Giessen) show that the description of the low-energy $E1$ strength strongly depends on the modeling of the underlying mean field.
The strength distributions are also sensitive to the interaction with multi-phonon states which redistribute the strength and can shift large parts in or out of the experimentally accessible energy region.
As highlighted by the present examples (Tab.~\ref{tab:Sumtheo}) one can only put a word of warning to attempts to establish systematics of the PDR by summing over more or less arbitrarily defined excitation energy regions 
It should be pointed out that all models discussed here predict an increase of the strength of the PDR in the stable even-mass tin isotopes with neutron excess.
However, the predictions of the energy region, where the PDR is confined, differ appreciably.
One possible explanation of the strong variations in Tab.~\ref{tab:Sumtheo} may be slight differences in the onset of the GDR strength. 

The experimental observable ($B(E1)$ strength) does not allow any conclusion on the nature of the excitation, thus no separation into PDR and GDR contributions is possible.  
Nevertheless, a linear increase of the summed strength with neutron excess is consistent with the data in $^{112,116,124}$Sn (cf.~Tab.~\ref{tab:Sum}) but clearly not for $^{120}$Sn.
While the absolute corrections necessary to estimate the full experimental $B(E1)$ strength are large, the methods to estimate their magnitude described above and below predict similar correction values for all four isotopes.
Thus, it is unlikely that the unexpected behavior of the low-energy $E1$ strength in $^{120}$Sn is caused by the experimental limitations. 

\subsection{Impact of the experimental sensitivity limit}

\begin{figure}[t]
    \includegraphics[angle=0,width=8.5cm]{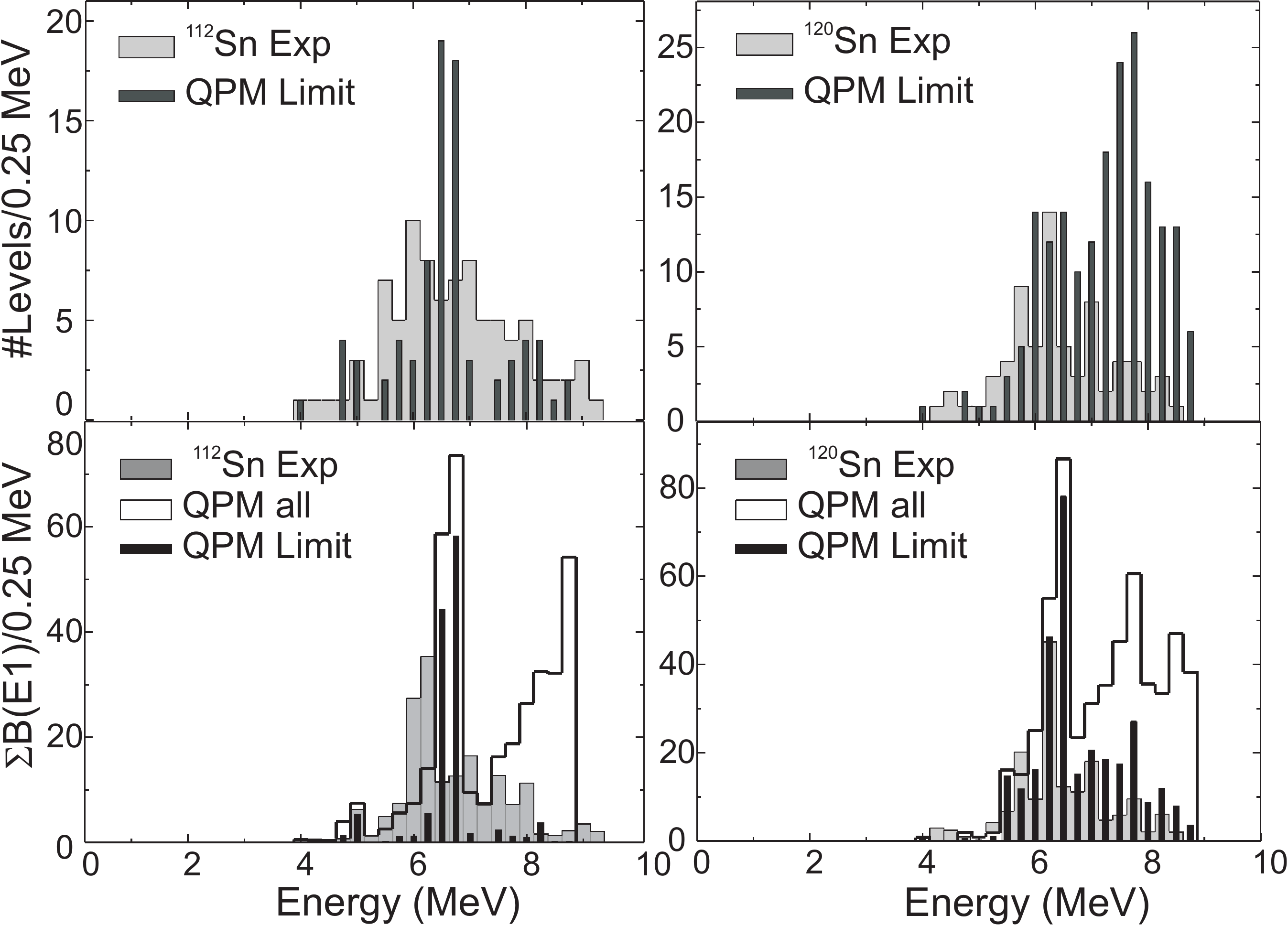}
        \caption{\label{fig:endist}
Distribution of the number of levels (top) and summed $B(E1)$ strengths (bottom) as a function of excitation energy in 250 keV bins for $^{112}$Sn (left) and $^{120}$Sn (right).
The experimental results (grey bars) are compared to the full QPM calculations (open bars) and taking into account the experimental sensitivity limits (black bars).}
\end{figure}

The experimental sensitivity limits given in Ref.~\cite{oze08} should be considered in the comparison with the theoretical strength distributions as has been discussed in Refs.~\cite{sav08,sav11}.
They represent the minimum $B(E1)$ strength of transitions which can be determined with at least $1 \sigma$  accuracy. 
The impact is illustrated in Fig.~\ref{fig:endist} for $^{112}$Sn and $^{120}$Sn, respectively.
The grey bars shows the number of experimentally observed levels (upper row) and the summed $B(E1)$ strength (lower row) as a function of excitation energy in bins of 250 keV.
They are compared to the QPM Darmstadt results neglecting (QPM all, open bars) or considering (QPM limit, black bars) the experimental sensitivity limit.

Taking into account the sensitivity limit, the obserevable $B(E1)$ strength above 6 MeV is significantly reduced.
Above 8 MeV in $^{112}$Sn and 7.5 MeV in $^{120}$Sn, the larger part of the strength stems from weak transitions which would be unobserved in the present experiments. 
For $^{120}$Sn, fairly good agreement with the experimental distribution is achieved after the correction, while for $^{112}$Sn the strength at higher excitation energies gets somewhat to small.
The numbers of excited levels per energy interval is fairly well reproduced but slightly too small for $^{112}$Sn, while for $^{120}$Sn they are still overpredicted at higher energy. 
Overall, as in a similar study of $N = 82$ isotones \cite{sav11}, the agreement between experiment and QPM predictions clearly improves with consideration of the experimental thresholds.

The integrated $B(E1)$ strengths of 0.192 e$^{2}$fm$^{2}$ ($^{112}$Sn) and 0.132 e$^{2}$fm$^{2}$ ($^{120}$Sn) in the QPM calculations below the sensitivity limits are remarkably close to the amounts of unresolved strengths deduced from the fluctuation analysis. 
We note that the agreement is not self-evident because of the different assumptions underlying both types of analysis.
Strong transitions as analyzed in Sec.~\ref{subsec:BE1} are little affected by admixtures of weak transitions due to the mechanism discussed  in Sec.~\ref{subsec:fluc-method}, while the mixing of transitions below the sensitivity limit may contribute to the fluctuations.
Therefore, we conclude that the good agreement is not fortituous but both methods allow an estimate of the quasi-continuum strength missed by a restriction of the data analysis to statistically significant peaks above background in the $(\gamma,\gamma')$ spectra.
Having two independent methods at hand is particularly helpful since their limits are defined by very different quantities. 
In case of the fluctuation analysis, there is an upper limit of the excitation energy defined by the statistics of the experimental spectra.
The extraction of the $B(E1)$ strength above the experimental threshold depends on the proper description of the fragmentation by the truncated theoretical model spaces (limited e.g.\ to $3p3h$ states in the present example). 

\section{Conclusions and Outlook}
\label{sec:conclusion}

This work presents high-resolution nuclear resonance fluorescence experiments on $^{112}$Sn and $^{120}$Sn to investigate the $E1$ response below the neutron threshold. 
The strongest transitions have been observed around the energy region of 6-7 MeV and the distributions of extracted $B(E1)$ transitions have a resonance-like structure, but in the case of $^{120}$Sn the strength is more fragmented. 
The summed $B(E1)$ strengths  are 0.175(24) e$^{2}$fm$^{2}$ and 0.164(31) e$^{2}$fm$^{2}$ with centroid energies 6.7 MeV and 6.6 MeV for $^{112}$Sn and $^{120}$Sn, respectively.

A fluctuation analysis has been applied to the ($\gamma$,$\gamma^{\prime}$) spectra to estimate the amount of unresolved strength which might be hidden in the background due to the finite energy resolution. 
The analysis is based on the autocorrelation function of the normalized spectrum which provides information about the mean level spacings in the investigated excitation energy interval.
These can be compared to model prediction of the nuclear level densities which allows to adjust the background. 
Three different theoretical models have been used, viz.\ BSFG1 \cite{rau97}, BSFG2 \cite{egi05} and HF-BCS \cite{dem01}. 
The backgrounds determined from these models produce a very similar amount of unresolved strength in the spectra and the differences between the models have been used to determine the uncertainties of this procedure. 
The fluctuation analysis shows that the amount of unresolved strength on ($\gamma$,$\gamma^{\prime}$) spectra is significant and the method should be applied to other data taken with the NRF method, at least for medium-heavy and heavy nuclei where $J^\pi = 1^-$ level densities in the energy region of the PDR typically fulfill Eq.~(\ref{eq:DelE}).

The measured $B(E1)$ strength distributions below the neutron threshold in $^{112}$Sn and $^{120}$Sn has been compared to microscopic model calculations. 
The QPM Darmstadt calculations reasonably reproduce the fragmentation of the transition strengths for $^{112}$Sn and $^{120}$Sn and also the total strength if one corrects for the experimentally unobservable part. 
The QPM Giessen and the RQTBA calculations predict much larger strengths than the experimental results if summed over the experimentally accessed energy region up to 9 MeV.
However, the agreement of the 2-phonon RQTBA predictions with the data is much better if the summation is restricted to an uper limit of 8 MeV.
Despite the problems discussed above to extract the full $E1$ strength from the $(\gamma,\gamma')$ data, the present results do not support an increase with neutron excess predicted by all models.
The modeling of the main corrections (branching ratios to excited states and unresolved strength) do not show systematic dependencies as a function of neutron number for $^{112-124}$Sn.  
The differences between the theoretical approaches indicate a strong sensitivity to the underlying mean field pointing towards a single-particle rather than a collective interpretation as discussed e.g.\ in Refs.~\cite{rei13,pap14,co13}.
  
For an improved understanding experimental information on the complete ground-state $E1$ response below and above the neutron threshold would be important, in particular to further investigate the unexpected behavior of $^{120}$Sn established in the present work. 
This is possible e.g.\ with high-resolution (p,p$^{\prime}$) experiments at 0$^{\circ}$ at RCNP Osaka university \cite{tam11,pol12}, with photon scattering coincidence experiments at HI$\gamma$S \cite{loh13} (below neutron threshold), and with the new NEPTUN tagger facility at the S-DALINAC \cite{sav10}. 
A corresponding ($p,p^{\prime}$) experiment was performed for the $^{120}$Sn case at RCNP \cite{kru10} and a study of $^{112,116,124}$Sn has been approved \cite{vnc14}.
Experimental studies of the $(\gamma,\gamma' \gamma'')$ and $(\gamma,n)$ reactions on $^{112,116,120,124}$Sn at the NEPTUN facility are underway.

\begin{acknowledgments}

R.~Eichhorn and the S-DALINAC crew are thanked for their effort in providing excellent beams and the GSI for the loan of the enriched $^{112}$Sn target.
We are grateful to V.~Yu.~Ponomarev for providing us with the results of his calculations and for important discussions.
We are indebed to F.~Siebenh\"uhner for his contribution to the analysis of the data.
B.\"{O}-T. acknowledges financial support from the DAAD sandwich program during her stay in Germany. 
This work was supported by the DFG under contract SFB 634, by the Alliance Program of the Helmholtz Association (HA216/EMMI), by the BMBF project 06GI9109, and by the US-NSF grant PHY-1204486 and the National Superconducting Cyclotron Laboratory at Michigan State University.

\end{acknowledgments}


\begin{thebibliography}{abc99x}
%\begin{thebibliography}{00}

\bibitem{sav13}
D. Savran, T. Aumann, and A. Zilges,
%
Prog. Part. Nucl. Phys. {\bf 70}, 210 (2013).

\bibitem{pie06}
J.~Piekarewicz, 
%
Phys.~ Rev. C {\bf 73}, 044325 (2006).

\bibitem{kli07}
A. Klimkiewicz {\it et al.},
%A.~Klimkiewicz, N.~Paar, P.~Adrich, M.~Fallot, K.~Boretzky, T.~Aumann,
%D.~Cortina~Gil, U.~Datta~Pramanik, Th.W.~Elze, H.~Emling, H.~Geissel,
%M.~Hellstr\"om, K.L.~Jones, J.V.~Kratz, R.~Kulessa, C.~Nociforo,
%R.~Palit, H.~Simon, G.~Surowka, K.~S\"ummerer, D.~Vretenar, and %W.~Walu,
%
Phys.~Rev. {\bf C76}, 051603(R) (2007).

\bibitem{tso08}
N. Tsoneva and H. Lenske,
%
Phys. Rev. C {\bf 77}, 024321 (2008).

\bibitem{pie11}
%
J. Piekarewicz,
%
Phys. Rev. C {\bf 83}, 034319 (2011).
 
\bibitem{ina11}
%
T. Inakura, T. Nakatsukasa, and K. Yabana,
%
Phys. Rev. C {\bf 84}, 021302 (2011).

\bibitem{rei10}
%
P.-G. Reinhard and W. Nazarewicz,
%
Phys. Rev. C {\bf 81}, 051303 (2010).

\bibitem{rei13}
%
P.-G. Reinhard and W. Nazarewicz,
%
Phys. Rev. C {\bf 87}, 014324 (2013).

\bibitem{car10}
%
A. Carbone, G. Col\`{o}, A. Bracco, L.-G. Cao, P. F. Bortignon, F.  Camera, and O. Wieland,
%
 Phys. Rev. C {\bf 81}, 041301 (2010).

\bibitem{fat12}
%
F. J. Fattoyev and J. Piekarewicz,
%
Phys. Rev. C {\bf 86}, 015802 (2012).
 
\bibitem{tsa12}
M. B. Tsang {\it et al.},
%M. B. Tsang, J. R. Stone, F. Camera, P. Danielewicz, S. Gandolfi, K. Hebeler, C. J. Horowitz, J. Lee, W. G. Lynch, Z. Kohley, R. Lemmon, P. Möller, T. Murakami, S. Riordan, X. Roca-Maza, F. Sammarruca, A. W. Steiner, I. Vida\~na, and S. J. Yennello,
%
Phys. Rev. C {\bf 86}, 015803 (2012). 

%\bibitem{gor98}
%
%S. Goriely,
%
%Phys. Lett. B {\bf 436}, 10 (1998).

\bibitem{gor04}
%
S. Goriely, E. Khan, and M. Samyn,
%
Nucl. Phys. {\bf A739}, 331 (2004).

\bibitem{lit09} 
%E. Litvinova, et al.,
E. Litvinova, H. P. Loens, K. Langanke, G. Mart\'{\i}nez-Pinedo, T. Rauscher, P. Ring, F.-K. Thielemann, and V. Tselyaev,
%
Nucl. Phys. {\bf A823}, 26 (2009).

\bibitem{dao12}
%
I. Daoutidis and S. Goriely,
%
Phys. Rev. C {\bf 86}, 034328 (2012).

\bibitem{lan71}
A.M. Lane,
%
Ann. Phys. (N.Y.) {\bf 63}, 171 (1971).

\bibitem{pap14}
P. Papakonstantinou, H. Hergert, V. Yu. Ponomarev, and R. Roth,
%
Phys. Rev. C {\bf 89}, 034306 (2014).

\bibitem{rye02}
N.~Ryezayeva, T.~Hartmann, Y.~Kalmykov, H.~Lenske, P.~von
Neumann-Cosel, V.~Yu.~Ponomarev, A.~Richter, A.~Shevchenko, S.~Volz, and J.~Wambach, 
%
Phys.~Rev.~Lett.~{\bf 89}, 272502  (2002).

\bibitem{ton10}
%A.P. Tonchev et al.,
A. P. Tonchev, S. L. Hammond, J. H. Kelley, E. Kwan, H. Lenske, G. Rusev, W. Tornow, and N. Tsoneva,
%
Phys. Rev. Lett. {\bf 104}, 072501 (2010).

\bibitem{lit10}
E. Litvinova, P. Ring, and V. Tselyaev,
%
Phys. Rev. Lett. {\bf 105}, 022502  (2010).

\bibitem{adr05}
P. Adrich {\it et al.},
%P.~Adrich, A.~Klimkiewicz, M.~Fallot, K.~Boretzky, T.~Aumann, D.~Cortina~Gil, U.~Datta~Pramanik, Th.~W.~Elze, H.~Emling, H.~Geissel, M.~Hellstr\"om, K. L.~Jones, J. V.~Kratz, R.~Kulessa, Y.~Leifels, C.~Nociforo, R.~Palit, H.~Simon, G.~Surowka, K.~S\"ummerer, and W.~Walu,
Phys.~Rev.~Lett. {\bf 95}, 132501 (2005).

\bibitem{wie09}
O. Wieland {\em et al.},
%O. Wieland, A. Bracco, F. Camera, G. Benzoni, N. Blasi, S. Brambilla, F. C. L. Crespi, S. Leoni, B. Million, R. Nicolini, A. Maj, P. Bednarczyk, J. Grebosz, M. Kmiecik, W. Meczynski, J. Styczen, T. Aumann, A. Banu, T. Beck, F. Becker, L. Caceres, P. Doornenbal, H. Emling, J. Gerl, H. Geissel, M. Gorska, O. Kavatsyuk, M. Kavatsyuk, I. Kojouharov, N. Kurz, R. Lozeva, N. Saito, T. Saito, H. Schaffner, H. J. Wollersheim, J. Jolie, P. Reiter, N. Warr, G. deAngelis, A. Gadea, D. Napoli, S. Lenzi, S. Lunardi,, D. Balabanski, G. LoBianco, C. Petrache, A. Saltarelli, M. Castoldi, A. Zucchiatti, J. Walker, and A. B�rger,
%
Phys. Rev. Lett. {\bf 102}, 092502 (2009).

\bibitem{ros13}
D. M. Rossi {\it et al.},
%D. M. Rossi,1,2,* P. Adrich,1 F. Aksouh,1,† H. Alvarez-Pol,3 T. Aumann,4,1,‡ J. Benlliure,3 M. Bo¨hmer,5 K. Boretzky,1 E. Casarejos,6 M. Chartier,7 A. Chatillon,1 D. Cortina-Gil,3 U. Datta Pramanik,8 H. Emling,1 O. Ershova,9 B. Fernandez-Dominguez,3,7 H. Geissel,1 M. Gorska,1 M. Heil,1 H. T. Johansson,10,1 A. Junghans,11 A. Kelic-Heil,1 O. Kiselev,1,2 A. Klimkiewicz,1,12 J.V. Kratz,2 R. Kru¨cken,5 N. Kurz,1 M. Labiche,13,14 T. Le Bleis,1,9,15 R. Lemmon,14 Yu.A. Litvinov,1 K. Mahata,1,16 P. Maierbeck,5 A. Movsesyan,4 T. Nilsson,10 C. Nociforo,1 R. Palit,17 S. Paschalis,4,7 R. Plag,9,1 R. Reifarth,9,1 D. Savran,18,19 H. Scheit,4 H. Simon,1 K. Su¨mmerer,1 A. Wagner,11 W. Walus´,12 H. Weick,1 and M. Winkler1
%
Phys. Rev. Lett. {\bf 111}, 242503 (2013).

\bibitem{kne96}
U.~Kneissl, H. H.~Pitz, and A.~Zilges,  
%
Prog. Part. Nucl. Phys. ~{\bf 37}, 349 (1996).

\bibitem{tam11}
A. Tamii {\em et al.},
%A. Tamii, I. Poltoratska, P. von Neumann-Cosel, Y. Fujita, T. Adachi, C.A. Bertulani, J. Carter, M. Dozono, H. Fujita, K. Fujita, K. Hatanaka, A.M. Heilmann, D. Ishikawa, M. Itoh, H. J. Ong, T. Kawabata, Y. Kalmykov, E. Litvinova, H. Matsubara, K. Nakanishi, R. Neveling, H. Okamura, B. \"Ozel-Tashenov, V.Yu. Ponomarev, A. Richter, B. Rubio, H. Sakaguchi, Y. Sakemi, Y. Sasamoto, Y. Shimbara, Y. Shimizu, F.D. Smit, %T. Suzuki, Y. Tameshige, J. Wambach, R. Yamada, M. Yosoi, and J. Zenihiro,
%
Phys. Rev. Lett. {\bf 107}, 062502 (2011).

\bibitem{pol92}
T. D. Poelhekken {\it et al.},
%T. D. Poelhekken, S. K. B. Hesmondhalgh, H. J. Hofmann, A. van der %Woude, and M. N. Harakeh,
Phys. Lett. {\bf B278}, 423 (1992).

\bibitem{sav06}
%D. Savran, et al.,
D. Savran, M. Babilon, A. M. van den Berg, M. N. Harakeh, J. Hasper, A. Matic, H. J. W\"ortche, and A. Zilges,
%
Phys. Rev. Lett. {\bf 97}, 172502 (2006).

\bibitem{end10}
J. Endres {\it et al.},
%J. Endres, E. Litvinova, D. Savran, P. A. Butler, M. N. Harakeh, S. Harissopulos, R.-D. Herzberg, R. Kr\"ucken, A. Lagoyannis, N. Pietralla, V. Yu. Ponomarev, L. Popescu, P. Ring, M. Scheck, K. Sonnabend, V. I. Stoica, H. J. W\"{o}rtche, and A. Zilges,
%
Phys. Rev. Lett. {\bf 105}, 212503 (2010).

\bibitem{paa09}
N. Paar, Y. F. Niu, D. Vretenar, and J. Meng,
%
Phys. Rev. Lett. {\bf 103}, 032502 (2009).

\bibitem{roc12}
%
X. Roca-Maza, G. Pozzi, M. Brenna, K. Mizuyama, and G. Col\`{o},
%
Phys. Rev. C {\bf 85}, 024601 (2012).

\bibitem{vre12}
%
D. Vretenar, Y. F. Niu, N. Paar, and  J.Meng,
%
Phys. Rev. C  {\bf 85}, 044317 (2012).

\bibitem{yuk12}
%
E. Y\"uksel, E. Khan, and K. Bozkurt,
%
Nucl. Phys. {\bf A877}, 35 (2012).

\bibitem{lan14}
E. G. Lanza, A. Vitturi, E. Litvinova, and D. Savran,
%
Phys. Rev. C {\bf 89}, 041601(R) (2014).

%\bibitem{moh71}
%R.~Mohan, M.~Danos, and L. C.~Biedenharn, Phys.~Rev.~{\bf C3}, 1740 (1971).

%\bibitem{bar72}
%G. A.~Bartholomew, E. D.~Earle, A. J.~Ferguson, J .W.~Knowles, and M. A.~Lone, Adv.~Nucl.~Phys.~{\bf 7}, 299 (1973).

\bibitem{har00}
T.~Hartmann, J.~Enders, P.~Mohr, K.~Vogt, S.~Volz, and A.~Zilges,
Phys.~Rev.~Lett.~{\bf 85}, 274 (2000).

\bibitem{isa11}
J. Isaak {\it et al.}, 
%J. Isaak, D. Savran, M.  Fritzsche, D.Galaviz, T. Hartmann, S.  Kamerdzhiev, J. H. Kelley, E.  Kwan, N.Pietralla, C. Romig, G. Rusev, K. Sonnabend, A. P. Tonchev, W. Tornow, and A. Zilges,
%
Phys. Rev. C {\bf 83}, 034304 (2011).

\bibitem{sch07}
R. Schwengner {\it et al.},
%R.~Schwengner, G.~Rusev, N.~Benouaret, R.~Beyer, M.~Erhard, E.~Grosse, A. R.~Junghans, J.~Klug, K.~Kosev, N.~Kostov, C.~Nair, N.~Nankov, K. D.~Schilling, and A.~Wagner, 
%
Phys.~Rev. ~{\bf C76}, 034321 (2007).

\bibitem{sch08}
R. Schwengner {\it et al.},
%R. Schwengner, G. Rusev, N. Tsoneva, N. Benouaret, R. Beyer, M. Erhard, E. Grosse, A. R. Junghans, J. Klug, K. Kosev, H. Lenske, C. Nair, K. D. Schilling, A. Wagner,
%
Phys.Rev. C {\bf 78}, 064314 (2008).

\bibitem{sch13}
R. Schwengner {\it et al.},
%R.Schwengner, R.Massarczyk, G.Rusev, N.Tsoneva, D.Bemmerer, R.Beyer, R.Hannaske, A.R.Junghans, J.H.Kelley, E.Kwan, H.Lenske, M.Marta, R.Raut, K.D.Schilling, A.Tonchev, W.Tornow, A.Wagner
Phys. Rev. C {\bf 87}, 024306 (2013).


\bibitem{gov98}
K.~Govaert, F.~Bauwens, J.~Bryssinck, D.~De~Frenne, E.~Jacobs, W.~Mondelaers, L. Govor, and V. Yu. Ponomarev, 
%
Phys.~Rev. ~{\bf C57}, 2229 (1998).

%\bibitem{end98}
%J. Enders {\it et al.},
%J.~Enders, P.~von Brentano, J.~Eberth, R.-D.~Herzberg, N.~Huxel, H.~Lenske, P.~von Neumann-Cosel, N.~Nicolay, N.~Pietralla, H.~Prade, J.~Reif, A.~Richter, C.~Schlegel, R.~Schwengner, S.~Skoda, H.G.~Thomas, I.~Wiedenh\"{o}ver, G.~Winter, and A.~Zilges, 
%Nucl.~Phys.~{\bf A636}, 139 (1998).

\bibitem{zil02}
A.~Zilges, S.~Volz, M.~Babilon, T.~Hartmann, P.~Mohr, and K.~Vogt,
%
Phys.~Lett.~B {\bf 542}, 43 (2002).

\bibitem{vol06}
S.~Volz, N.~Tsoneva, M.~Babilon, M.~Elvers, J.~Hasper, R.-D.~Herzberg,
H.~Lenske, K.~Lindenberg, D.~Savran, and A.~Zilges, 
%
Nucl.~Phys. ~{\bf A779}, 1 (2006).

\bibitem{sav08}
%D. Savran {\em et al.},
D. Savran, M. Fritzsche, J. Hasper, K. Lindenberg, S. M\"uller, V. Yu. Ponomarev, K. Sonnabend, and A. Zilges,
%
Phys. Rev. Lett. {\bf 100}, 232501 (2008).

\bibitem{end00}
J. Enders {\it et al.},
%
%J. Enders, P. von Brentano, J. Eberth, A. Fitzler, C. Fransen, R.-D. Herzberg, H. Kaiser, L. K\''aubler, P. von Neumann-Cosel, N. Pietralla, V. Yu. Ponomarev, H. Prade, A. Richter, H. Schnare, R. Schwengner, S. Skoda, H. G. Thomas, H. Tiesler, D. Weisshaar, and I. Wiedenh\''over,
%
Phys.~Lett. B {\bf 486}, 279 (2000).

\bibitem{end03}
J. Enders {\it et al.},
%J.~Enders, P.~von Brentano, J.~Eberth, A.~Fitzler, C.~Fransen, R.-D.~Herzberg, H.~Kaiser, L.~K\"{a}ubler, P.~von Neumann-Cosel, N.~Pietralla, V.~Yu.~Ponomarev, A.~Richter, R.~Schwengner, and I.~Wiedenh\"{o}ver,
Nucl.~ Phys.~{\bf A724}, 243 (2003).

\bibitem{sch10}
R. Schwengner {\it et al.},
%R. Schwengner, R. Massarczyk, B.A. Brown, R. Beyer, F. D\"{o}nau, M. Erhard, E. Grosse, A.R. Junghans, K. Kosev, C. Nair, G. Rusev, K.D. Schilling, and A. Wagner,
%
Phys. Rev. C {\bf 81}, 054315 (2010).

\bibitem{paa07}
%N. Paar {\em et al.},
N. Paar, D. Vretenar, E. Khan, and G. Col\`{o},
%
Rep. Prog. Phys. {\bf 70}, 691 (2007).

\bibitem{aum12}
GSI Experiment S412, spokespersons T. Aumann and K. Boretzky.

\bibitem{sar04}
D. Sarchi, P. F. Bortignon, and G. Col\`o,
%
Phys. Lett. B {\bf 601}, 24 (2004).

\bibitem{tso04a}
N. Tsoneva, H. Lenske, and Ch. Stoyanov,
%
Nucl. Phys. {\bf A731}, 273 (2004).

\bibitem{tso04b}
N. Tsoneva, H. Lenske, and Ch. Stoyanov,
%
Phys. Lett. B {\bf 586}, 213 (2004.

%
\bibitem{vre04}
D. Vretenar, T. Nik\v si\'c, N. Paar, and P. Ring,
%
Nucl. Phys. {\bf A731}, 281 (2004).

%
\bibitem{paa05}
N. Paar, T. Nik\v si\'c, D. Vretenar, and P. Ring,
%
Phys. Lett. B {\bf 606}, 288 (2005).

%
\bibitem{kam06}
S.P. Kamerdzhiev,
%
Phys. At. Nucl. {\bf 69}, 1110 (2006).

\bibitem{ter06}
J. Terasaki and J. Engel,
%
Phys. Rev. C {\bf 74}, 044301 (2006).

\bibitem{lit08}
E.~Litvinova, P.~Ring, and V.~Tselyaev, 
%
Phys.~Rev. C{\bf 78}, 014312 (2008).

\bibitem{lit13}
E.~Litvinova, P.~Ring, and V.~Tselyaev, 
%
Phys.~Rev. C{\bf 88}, 044320 (2013).

\bibitem{lan09}
%E.G. Lanza, et al.,
E. G. Lanza, F. Catara, D. Gambacurta, M. V. Andres, and Ph. Chomaz,
%
Phys. Rev. C {\bf 79}, 054615 (2009).

\bibitem{co09} 
%G. Co', et al.,
G. Co', V. De Donno, C. Maieron, M. Anguiano, and A. M. Lallena,
%
Phys. Rev. C {\bf 80}, 014308 (2009); publishers note C {\bf 80}, 019910 (2009).

\bibitem{avd11}
%
A. Avdeenkov, S. Goriely, S. Kamerdzhiev, and S. Krewald,
%
Phys. Rev. C {\bf 83}, 064316 (2011).

\bibitem{dao11}
%
I. Daoutidis and P. Ring,
%
 Phys. Rev. C {\bf 83}, 044303 (2011).


\bibitem{rus08}
G. Rusev {\em et al.},
%G. Rusev, R. Schwengner, F. D\"{o}nau, M. Erhard, E. Grosse, A. R. Junghans, K. Kosev, K. D. Schilling, A. Wagner, F. Be\c{c}var, and M. Krticka,
%
Phys. Rev. C {\bf 77}, 064321 (2008).

\bibitem{oze08}
B. \"Ozel, PhD thesis, University of Cukurova (2008).

\bibitem{moh99}
P. Mohr, J. Enders, T. Hartmann, H. Kaiser, D. Schiesser, S. Schmitt, S. Volz, F. Wissel, and A. Zilges, 
%
Nucl. Instrum. Methods Phys. Res., Sect. A {\bf 423}, 480 (1999).

\bibitem{nndc}
NNDC, http://www.nndc.bnl.gov/.

\bibitem{bry99}
J. Bryssinck {\it et al.},
%J.~Bryssinck, L.~Govor, D.~Belic, F.~Bauwens, O.~Beck, P.~von Brentano, D.~De~Frenne, T.~Eckert, C.~Fransen, K.~Govaert, R.-D.~Herzberg, E.~Jacobs, U.~Kneissl, H.~Maser, A.~Nord, N.~Pietralla, H H.~Pitz, V.~Yu.~Ponomarev, and V.~Werner, 
Phys.~Rev. ~{\bf C59}, 1930 (1999).

\bibitem{pys06}
I. Pysmenetska {\it et al.},
%I.~Pysmenetska, S.~Walter, J.~Enders, H.~von~Garrel, O.~Karg, U.~Kneissl, C.~Kohstall, P.~von Neumann-Cosel, H H.~Pitz, V.~Yu.~Ponomarev, M.~Scheck, F.~Stedile, and S.~Volz, 
Phys.~Rev.~{\bf C73}, 017302 (2006).

\bibitem{sie06}
F.~Siebenh\"uhner,
%
B.Sc.\ thesis, Technische Universit\"{a}t Darmstadt (2005), unpublished.

\bibitem{hut02}
C. Hutter {\it et al.},
%C. Hutter, M. Babilon, W. Bayer, D. Galaviz, T. Hartmann, P. Mohr, S. Müller, W. Rochow, D. Savran, K. Sonnabend, K. Vogt, S. Volz, and A. Zilges,
%
Nucl. Instrum. Methods Phys. Res., Sect. A {\bf 489}, 247 (2002).

\bibitem{isa13}
J. Isaak {\it et al.}, 
%J. Isaak, D. Savran, M. Krticka, M. W. Ahmed, J. Beller, E. Fiori, J. Glorius, J. H. Kelley, B. L\''oher, N. Pietralla, C. Romig, G. Rusev, M. Scheck, L. Schnorrenberger, J. Silva, K. Sonnabend, A. P. Tonchev, W. Tornow, H. R. Weller, and M. Zweidinger,
%
Phys.Lett. B {\bf 727}, 361 (2013).

\bibitem{hey10}
K. Heyde, P. von Neumann-Cosel, and A. Richter,
%
Rev. Mod. Phys. {\bf 82}, 2365 (2010).

\bibitem{pol05}
I.~Poltaratska, 
%
Diploma thesis, Technische Universit\"{a}t Darmstadt (2005), unpublished.

\bibitem{ripl}
R. Capote {\it et al.}, 
%
Nucl. Data Sheets {\bf 110}, 3107 (2009).

\bibitem{rau97}
T. Rauscher, F.-K. Thielemann, and K.-L. Kratz, 
%
Phys. Rev. C {\bf 56}, 1613 (1997).

\bibitem{egi05}
T. von Egidy and D. Bucurescu,
%
Phys. Rev. C {\bf 72}, 044311 (2005); erratum C {\bf 73}, 049901
(2006).

\bibitem{uts06}
H. Utsunomiya, P. Mohr, A. Zilges, and M. Rayet,
%
Nucl. Phys. {\bf A777}, 459 (2006).

\bibitem{kal07}
Y.~Kalmykov, C.~\"Ozen, K.~Langanke, G.~Mart\'{\i}nez-Pinedo, P.~von
Neumann-Cosel, and A.~Richter, 
%
Phys. Rev. Lett. {\bf 99}, 202502 (2007).

\bibitem{loh13}
B. L\"oher {\it et al.}, 
%B. L\''oher, V. Derya, T. Aumann, J. Beller, N. Cooper, M. Duchene, J. Endres, E. Fiori, J. Isaak,J. Kelley, M. Kn\''orzer,  N. Pietralla, C. Romig, D. Savran, M. Scheck, H. Scheit,J. Silva, A. Tonchev, W. Tornow, H. Weller, V. Werner, and A. Zilges,
%
Nucl. Instrum. Methods Phys. Res., Sec. A {\bf 723}, 136 (2013).

\bibitem{end97}
J.~Enders, N.~Huxel, P.~von Neumann-Cosel, and A.~Richter,
%
Phys. Rev. Lett. {\bf 79}, 2010 (1997).

\bibitem{end98}
J. Enders, N. Huxel, U. Kneissl, P. von Neumann-Cosel, H. H. Pitz, and A. Richter,
%
Phys. Rev. C {\bf 57}, 996 (1998).

\bibitem{hux99}
N. Huxel {\it et al.},
%N.~Huxel, P.~von Brentano, J.~Eberth, J.~Enders, R.-D.~Herzberg, P.~von Neumann-Cosel, N.~Nicolay, N.~Pietralla, H.~Prade, C.~Rangacharyulu, J.~Reif, A.~Richter, C.~Schlegel, R.~Schwengner, S.~Skoda, H. G.~Thomas, I.~Wiedenh\"{o}ver, G.~Winter, and A.~Zilges, 
Nucl.~Phys.~{\bf A645}, 239 (1999).

\bibitem{nor03}
A. Nord {\it et al.},
%A. Nord, J. Enders, A. E. de Almeida Pinto, D. Belic, P. von Brentano, C. Fransen, U. Kneissl, C. Kohstall, A. Linnemann, P. von Neumann-Cosel, N. Pietralla, H. H. Pitz, A. Richter, F. Stedile, and V. Werner,
%
Phys. Rev. C {\bf 67}, 034307 (2003).

\bibitem{jon76}
B. Jonson {\it et al.}, 
Proceedings of the 3rd International Conference on
Nuclei far from Stability, Cargese, 1976, CERN Report No. 76-13 (1976)
p. 277.

\bibitem{han79}
P. G.~Hansen, 
%
Annu. Rev. Nucl. Part. Sci. ~{\bf 29}, 69 (1979).

\bibitem{han90}
P. G. Hansen, B. Jonson, and A. Richter,
%
Nucl. Phys. {\bf A518}, 13 (1990).

\bibitem{end04}
J. Enders, T. Guhr, A. Heine, P. von~Neumann-Cosel, V. Yu. Ponomarev,  A. Richter, and J. Wambach,
%
Nucl. Phys. {\bf A741}, 3 (2004).

\bibitem{kal06}
Y. Kalmykov {\it et al.},
%
Phys. Rev. Lett. {\bf 96}, 012502 (2006)

\bibitem{usm11}
I. Usman {\it et al.},
%
Phys. Rev. C {\bf 84}, 054322 (2011).

\bibitem{pol14}
I. Poltoratska, R. W. Fearick, A. M. Krumbholz, E. Litvinova, H. Matsubara, P. von Neumann-Cosel, V. Yu. Ponomarev, A. Richter, and A. Tamii,
%
Phys. Rev. C (to be published); arXiv:1403.7652.

\bibitem{she08}
A. Shevchenko {\it et al.},
%
Phys. Rev. C {\bf 77}, 024302 (2008).

\bibitem{dem01}
P.~Demetriou and S.~Goriely, 
%
Nucl. Phys. {\bf A695}, 95 (2001).

\bibitem{has07}
J. Hasper, private communication.

\bibitem{sol00}
V. G. Soloviev, {\it Theory of Atomic Nuclei: Quasiparticles and Phonons} (Institute of Physics, Bristol, 1992).

\bibitem{pon79}
V. Yu. Ponomarev, V. G. Soloviev, Ch. Stoyanov, and A. I. Vdovin,
%
Nucl. Phys. {\bf  A323}, 446 (1979).

\bibitem{pon99}
V. Yu. Ponomarev and P. von Neumann-Cosel,
%
Phys. Rev. Lett. {\bf 82}, 501 (1999).

\bibitem{vnc99}
P. von Neumann-Cosel {\it et al.},
%P. von Neumann-Cosel, F. Neumeyer, S. Nishizaki, V. Yu. Ponomarev, C. Rangacharyulu, B. Reitz, A. Richter, G. Schrieder, D. I. Sober, T.Waindzoch, and J.Wambach, 
%
Phys. Rev. Lett. {\bf 82}, 1105 (1999).

\bibitem{she04}
A. Shevchenko {\em et al.},
%A. Shevchenko, J. Carter, R. W. Fearick, S. V. F\"ortsch, H. Fujita, Y. Fujita, Y. Kalmykov, D. Lacroix, J. J. Lawrie, P. von Neumann-Cosel, R. Neveling, V. Yu. Ponomarev, A. Richter, E. Sideras-Haddad, F. D. Smit, and J. Wambach,
%
Phys. Rev. Lett. {\bf 93}, 122501 (2004).

\bibitem{wal11}
C. Walz, H. Fujita, A. Krugmann, P. von Neumann-Cosel, N. Pietralla, V. Yu. Ponomarev, A. Scheikh-Obeid, and J. Wambach, 
%
Phys. Rev. Lett. {\bf 106}, 062501 (2011).

\bibitem{rus13}
G. Rusev {\it et al.},
%G. Rusev, N. Tsoneva, F. D\''onau, S. Frauendorf, R. Schwengner, A. P. Tonchev, A. S. Adekola, S. L. Hammond, J. H. Kelley, E. Kwan, H. Lenske, W. Tornow, and A. Wagner,
%
Phys. Rev. Lett. {\bf 110}, 022503 (2013).

\bibitem{lit07}
E.~Litvinova, P.~Ring, and V.~Tselyaev, 
%
Phys. Rev. C {\bf 75}, 064308 (2007).

\bibitem{sav11}
D. Savran {\it et al.}
%D.Savran, M.Elvers, J.Endres, M.Fritzsche, B.Loher, N.Pietralla, V.Yu.Ponomarev, C.Romig, L.Schnorrenberger, K.Sonnabend, A.Zilges,
%
Phys. Rev. C {\bf 84}, 024326 (2011).

%\bibitem{tam09}
%A. Tamii {\em et al.},
%A.~Tamii, Y.~Fujita, H.~Matsubara, T.~Adachi, J.~Carter, M.~Dozono,
%H.~Fujita, K.~Fujita, H.~Hashimoto, K.~Hatanaka, T.~Itahashi,
%M.~Itoh, T.~Kawabata, K.~Nakanishi, S.~Ninomiya, A.B.~Perez-Cerdan,
%L.~Popescu, B.~Rubio, T.~Saito, H.~Sakaguchi, Y.~Sakemi,
%Y.~Sasamoto, Y.~Shimbara, Y.~Shimizu, F.D.~Smit, Y.~Tameshige,
%M.~Yosoi, and J.~Zenhiro,
%
%Nucl. Instrum. Methods Phys. Res., Sect. A {\bf 605}, 3 (2009).

%\bibitem{lin07}
%K.~Lindenberg, Dissertation D17, TU Darmstadt (2007).

\bibitem{co13}
G. Co', V. De Donno, M. Anguiano, and A. M. Lallena,
%
Phys. Rev. C {\bf 87}, 034305 (2013).

\bibitem{pol12}
I. Poltoratska {\it et al.},
%
Phys. Rev. C {\bf 85}, 041304(R) (2012).

\bibitem{sav10}
D. Savran {\it et al.},
%D. Savran, K. Lindenberg, J. Glorius, B. L\''oher, S. M\''uller, N. Pietralla, L. Schnorrenberger, V. Simon, K. Sonnabend, C. W\''alzlein, M. Elvers, J. Endres, J. Hasper, and A. Zilges,
%
Nucl. Instrum. Methods Phys. Res., Sec. A {\bf 613}, 232 (2010).

\bibitem{kru10}
A. M. Heilmann {\it et al.},
%
J. Phys.: Conf. Series {\bf 312}, 092029 (2011).

\bibitem{vnc14}
P. von Neumann-Cosel, A. Tamii, RCNP proposal E422.

%\end{thebibliography}
\end{thebibliography}
\end{document}